\documentclass[12pt]{article}

\usepackage{adjustbox}
\usepackage{amsfonts,amsmath,amssymb, amsthm}
\allowdisplaybreaks
\usepackage{appendix}
\usepackage{bm}
\usepackage{braket}
\usepackage[table]{xcolor}
    \definecolor{darkgreen}{rgb}{0,0.5,0}
    \definecolor{darkred}{rgb}{0.5,0,0}
    \definecolor{darkblue}{rgb}{0,0,0.6}
    \definecolor{purple}{rgb}{0.4,.2,0.7}
\usepackage{comment}
\usepackage{esint}
\usepackage{euscript}
\usepackage{float}
\usepackage{graphicx}
\usepackage[numbers,sort&compress]{natbib}
\usepackage[hyperfootnotes = true, colorlinks = true, linkcolor = darkblue, citecolor = purple]{hyperref}
\usepackage{import}
\usepackage{makecell}
\usepackage{multirow}
\usepackage{subcaption}
\usepackage{tcolorbox}
\usepackage{tikz}
\usepackage[normalem]{ulem}

\numberwithin{equation}{section}

\usepackage[margin = 2.2cm]{geometry}
\usepackage[ragged]{footmisc}
\newcommand{\Sym}{\mathrm{Sym}}
\newcommand{\dif}{\mathrm{d}}
\renewcommand{\vec}[1]{\bm{#1}}

\def\ZZZ{{\hbox{ Z\kern-1.6mm Z}}}
\def\RRR{{\hbox{ R\kern-2.4mm R}}}
\def\CCC{{\hbox{ C\kern-2.0mm C}}}
\def\zzz{{\hbox{z\kern-1mm z}}}

\def\ZZZ{\mathbb{Z}}
\def\RRR{\mathbb{R}}

\newcommand{\qeq}{{\hbox{=\kern-2.3mm ? \kern.5mm }}}
\renewcommand{\qeq}{=}

\newcommand{\eps}{\epsilon}

\newcommand{\OO}{{\cal O}}

\newcommand{\EE}{{\cal E}}

\newcommand{\wt}{\widetilde}

\newcommand{\be}{\begin{equation}}
\newcommand{\ee}{\end{equation}}
\newcommand{\ben}{\begin{eqnarray}\displaystyle}
\newcommand{\een}{\end{eqnarray}}

\newcommand{\refb}[1]{(\ref{#1})}
\newcommand{\p}{\partial}
\newcommand{\sectiono}[1]{\section{#1}\setcounter{equation}{0}}

\def\one{{\hbox{ 1\kern-.8mm l}}}
\def\zero{{\hbox{ 0\kern-1.5mm 0}}}

\newcommand{\bea}[1]{\begin{eqnarray}\label{#1} }
\newcommand{\eea}{\end{eqnarray}}

\renewcommand{\eqref}{\refb}

\usepackage{bm}

\begin{document}

\thispagestyle{empty}
\begin{center}
    ~\vspace{5mm}

    {\Large \bf

        All-Orders Classical Soft Graviton Theorems in the Newtonian Limit
        and Beyond
    }

    \vspace{0.4in}

    {\bf Debanjan Karan,$^1$ Babli Khatun,$^1$ Raghu Mahajan,$^1$\vskip1ex Biswajit Sahoo,$^{2}$ and Ashoke Sen$^{1}$}

    \vspace{0.4in}

    $^1$ International Centre for Theoretical Sciences, Bengaluru, Karnataka 560089, India \vskip1ex
    $^2$ School of Physical Sciences, Indian Association for the
Cultivation of Science (IACS),
2A \& 2B Raja S.C. Mullick Road, Kolkata 700032, India

    \vspace{0.1in}

    {\tt debanjan.karan@icts.res.in, babli.khatun@icts.res.in, raghu.mahajan@icts.res.in, spsbs@iacs.res.in, ashoke.sen@icts.res.in}
\end{center}

\vspace{0.4in}

\begin{abstract}

The classical soft graviton theorem aims to determine certain terms in the early- and
late-time tails of the gravitational waveform produced in a scattering process, using
only the asymptotic momenta of the scattered objects.
In this paper we find an infinite series of leading-logarithmic tail terms in the
Newtonian limit, where the incoming and outgoing particles move slowly, but the
scattering itself can be violent.
The all-orders conjecture of Alessio, Di Vecchia and Heissenberg agrees with our results
for two-particle scattering, but does not hold in general when the initial or final state
contains three or more particles.
We also propose an all-orders conjecture for these tails beyond the Newtonian limit,
but do not prove it. Nevertheless, the conjecture passes several nontrivial
consistency checks.

\end{abstract}

\pagebreak

\tableofcontents

\pagebreak

\sectiono{Introduction and summary} \label{s1}

The classical soft graviton theorem is the statement that, in a scattering process,
certain terms in the early-time growth and the late-time decay of the gravitational
waveform are determined by the momenta of the incoming and outgoing objects and the
position of the detector, without any knowledge of the interactions that take place
during the scattering \cite{1804.09193,1806.01872,1808.03288,1912.06413,2008.04376,
2105.08739,2106.10741,2408.08851}.
The relevant terms are those of the form $u^{-k}(\ln|u|)^{k-1}$, where $u$ is the
retarded time at the detector. This is not a theorem yet, in the sense that it has not
been proven for general values of $k$; it has been established only for $k\le 2$.
Our eventual goal is to prove the theorem and find explicit expressions for the
coefficients of these terms. In this paper we achieve this in the Newtonian limit,
where the incoming and outgoing particles are nonrelativistic, but still without
imposing any restriction on the interactions that take place during the scattering.
In particular, only the asymptotic velocities need to be small; at intermediate
stages of the scattering the particles can move relativistically.
We ignore long-range electromagnetic interactions in our analysis, but they can
easily be included if the incoming or outgoing objects are charged.

The soft graviton theorem originally arose as a statement about scattering amplitudes
involving low-frequency gravitons \cite{weinberg1,weinberg2}. The subject has a long
history, and we shall refrain from giving an overview of it. Our focus will be on four
spacetime dimensions, where the soft theorems are modified by infrared divergences and
acquire terms logarithmic in the frequency in the low-frequency
expansion \cite{1804.09193,1808.03288,1405.1015,2308.16807}.\footnote{The logarithmic
soft photon theorem and the associated waveform tails have also been derived in higher
dimensions~\cite{Sahoo:2026ulx}.}
In an appropriate classical limit, these can be translated into statements about the
gravitational waveform at low frequency, which in turn translate into statements about
the waveform at early and late retarded times. Once this connection is understood,
however, we can set aside the quantum soft theorem and study the classical soft graviton
theorem directly, by analyzing the classical gravitational waveform at large retarded
times or, equivalently, at low frequencies. This approach has been quite successful in
establishing the classical soft graviton theorem for $k\le 2$ \cite{1804.09193,1806.01872,
1808.03288,1912.06413,2008.04376,2105.08739,2106.10741,2308.16807}, and this is the
approach we shall follow in this paper.

The predictions of the soft graviton theorem have also been verified independently.
In particular, they have been tested against various other computations of gravitational
waves that do not assume low frequencies but from which the low-frequency limit can be
extracted \cite{Damour:1985cm,Blanchet:1992br,0501041,1310.1528,1409.4555,1812.08137,1901.10986,
2007.02077,DiVecchia:2023frv,2211.13120,2309.14925,2312.07452,2402.06361,
2402.06604,2402.06533,2407.02076,Aoude:2023dui,Georgoudis:2025vkk,Akhtar:2026pkm}.
Soft graviton theorems have been related to asymptotic symmetries, which can be studied
in their own right. For example, the leading soft graviton theorem determines the
memory, i.e., the permanent change in the waveform between early and
late times~\cite{mem1,mem2,mem3,mem4}, and is related to BMS
supertranslations~\cite{1411.5745}. Subleading soft theorems have also been related to
asymptotic symmetries in \cite{1903.09133,1912.10229,Compere:2025tzr,2205.11477,
2309.11220,2403.13053,2407.07978,DeAngelis:2025vlf,
Compere:2026jmk,2603.09844,2603.22681,2512.21239,
Banerjee:2026keq,2601.05951,2609.39253}. Finally, some unusual properties of the results
of the classical soft graviton theorem have been given an independent explanation in
\cite{2508.17919}.

These successes encourage us to extend the results to terms of the form
$u^{-k}(\ln|u|)^{k-1}$ for $k>2$. An all-orders structure for the classical soft
graviton theorem was proposed in~\cite{2008.04376}, although certain gauge-invariant
contributions were left undetermined. Subsequently, Alessio, Di Vecchia and Heissenberg
conjectured an explicit all-orders formula for gravity and checked it for two-body
scattering at Newtonian order~\cite{2407.04128}. A further test of this formula was
carried out in the probe approximation in~\cite{2408.07329}. For electromagnetic
scattering with gravitational interactions neglected, the corresponding coefficients
have been determined for an arbitrary number of particles~\cite{2501.07328}. The
electromagnetic two-particle result admits a simple resummation, producing a formula similar to the
one conjectured in~\cite{2407.04128}.
For three or more particles, however, the electromagnetic waveform has a
considerably more complicated structure than the corresponding formula
of~\cite{2407.04128}, which made it unlikely that the gravitational conjecture
of~\cite{2407.04128} holds for multiparticle scattering~\cite{2501.07328}.
This motivates a direct computation of the coefficients of
$u^{-k}(\ln|u|)^{k-1}$ in the gravitational waveform for scattering with an arbitrary
number of particles.

In this paper, we carry out this computation at leading Newtonian order.
We now describe the setup and summarize our results. We consider a scattering process in which incoming
particles of masses $m'_a$ and asymptotic velocities $\alpha\vec v'_a$ collide and
disperse into outgoing particles of masses $m_a$ and asymptotic velocities
$\alpha\vec v_a$, with $\alpha\ll1$. The interactions during the collision are left
unrestricted, while Newtonian gravity describes the long-range interaction between the
particles at early and late times.
The masses and rescaled velocities are held fixed as $\alpha\to0$, with nonzero
relative velocities for distinct particles in each asymptotic state.
We also assume that the energy carried away by any gravitational radiation
produced during the collision is small compared with the kinetic energies of the massive
particles in the initial and final states, which are of order $\alpha^2m$. This ensures
that the nonlinear memory sourced by this radiation, which we do not include, is
negligible compared with the $O(\alpha^2)$ waveform computed here.

The waveform is described by the trace-reversed metric fluctuation $e^{\mu\nu}$,
defined by $g_{\mu\nu}=\eta_{\mu\nu}+2h_{\mu\nu}$ and
$e_{\mu\nu}:=h_{\mu\nu}-\frac12\eta_{\mu\nu}h^\rho{}_{\rho}$, with
$\eta_{\mu\nu}=\mbox{diag}(-1,+1,+1,+1)$. For a detector at distance $R$,
the calculation retains the $1/R$ part of the waveform. The result is expressed in
terms of the rescaled retarded time $w:=\alpha^3u$, where $u$ is the retarded time at
the detector. At leading order, $O(\alpha^2)$, the calculation determines the memory
term and the highest logarithm at each inverse-time order, $w^{-k}(\ln|w|)^{k-1}$ for
$k\ge1$; lower logarithms and higher orders in $\alpha$ are omitted.
This selects coefficients of the late-time expansion rather than truncating it by
magnitude: for example, $w^{-3}(\ln w)^2$ is retained while $w^{-2}$ is not.
In this approximation, the late-time spatial waveform is
\ben \label{eij_late_time}
e^{ij}(w) &=& {G\alpha^2\over  R} \sum_{a} m_a  \Bigg[ 2 v_a^i v_a^j
+ 2 w^{-1} \, v_a^i \Bigg(c_a^j - \sum_{r\ge 1} d^j_{r,a} r z^r\Bigg)
+ 2 w^{-1} \, v_a^j \Bigg(c_a^i - \sum_{r\ge 1} d^i_{r,a} r z^r\Bigg) \nonumber \\
&+& w^{-1} \Bigg(v_a^i + c_a^i \, z\Bigg) \Bigg(  -c_a^j + \sum_{r\ge 1} d_{r,a}^j r(r+1)
z^r\Bigg) \nonumber \\
&+& w^{-1} \Bigg(v_a^j + c_a^j \, z\Bigg) \Bigg(  -c_a^i + \sum_{r\ge 1} d_{r,a}^i r(r+1)
z^r\Bigg)\Bigg],\qquad z:= \frac{\ln w}{w},\quad w\rightarrow \infty\, .
\een
Here $G$ is Newton's constant, $i,j$ are spatial indices, and the sum
runs over the outgoing particles. If we write the position of the $a$'th particle as a
function of rescaled time, $\vec X_a(w):=\vec X_a(u=w/\alpha^3)$, the vectors
$\vec c_a$ and $\vec d_{r,a}$ are the coefficients of $\ln w$ and $(\ln w/w)^r$,
respectively, in the asymptotic expansion of $\alpha^2\vec X_a(w)$; see \refb{e1.20}. Their expressions
in terms of the asymptotic masses and velocities are
\begin{align}
\vec c_a &= G\, \sum_{b\ne a}  m_b\, {\vec v_a -\vec v_b\over |\vec v_a-\vec v_b|^3}\, ,
\label{c_a}\\[2mm]
\vec d_{r,a} &= -{G\over 2^r r(r+1)}\sum_{b\ne a}
{m_b\over |\vec v_a-\vec v_b|^3}
\Bigg[ (\vec v_a-\vec v_b)
\sum_{\ell=[(r+1)/2]}^{r}
{(-1)^\ell(2\ell+1)!\over \ell!\,(2\ell-r)!\,(r-\ell)!}
\notag\\*
&\hskip 1.2in\times
{\bigl[(\vec v_a-\vec v_b)\cdot(\vec c_a-\vec c_b)\bigr]^{2\ell-r}
|\vec c_a-\vec c_b|^{2(r-\ell)}\over |\vec v_a-\vec v_b|^{2\ell}}
\notag\\*
&\hskip 0.9in+2(\vec c_a-\vec c_b)
\sum_{\ell=[r/2]}^{r-1}
{(-1)^\ell(2\ell+1)!\over \ell!\,(2\ell+1-r)!\,(r-\ell-1)!}
\notag\\*
&\hskip 1.2in\times
{\bigl[(\vec v_a-\vec v_b)\cdot(\vec c_a-\vec c_b)\bigr]^{2\ell+1-r}
|\vec c_a-\vec c_b|^{2(r-\ell-1)}\over |\vec v_a-\vec v_b|^{2\ell}}
\Bigg],\qquad r\ge1\, ,
\label{d_ra}
\end{align}
where $[s]$ denotes the largest integer not exceeding $s$. Once \refb{c_a} is known,
\refb{d_ra} determines each higher trajectory coefficient directly. Together with
energy-momentum conservation, the asymptotic Newtonian trajectories determine
\refb{eij_late_time} at the retained order, including both matter and gravitational
contributions. The retained coefficients depend only on the asymptotic masses and
velocities, and are otherwise independent of the details of the collision.

Equation~\refb{eij_late_time} gives only the spatial components of the waveform, and
was derived in harmonic gauge. If the detector lies in the direction of the spatial unit
vector $\vec n$, and $n=(1,\vec n)$ is the four-vector along the spacetime direction of
the outgoing graviton, then the harmonic gauge condition gives $n^\mu e_{\mu\nu}=0$ for the time-dependent part of the waveform.
This can be used to determine the time components of $e_{\mu\nu}$. Moreover, in the
final result we are free to transform the waveform by a gauge transformation
$h_{\mu\nu}\to h_{\mu\nu}+n_\mu\xi_\nu+n_\nu\xi_\mu$ for any four-vector $\xi$. Using
this freedom, we can transform $h_{ij}$ into a transverse-traceless tensor. The physical
content of \refb{eij_late_time} is therefore its transverse-traceless part.

At early times ($w\to-\infty$), the incoming coefficient $\vec c'_a$ follows
from \refb{c_a} by replacing $m_a,\vec v_a$ by $m'_a,\vec v'_a$ and reversing the
overall sign. Similarly, $\vec d'_{r,a}$ follows from \refb{d_ra} by replacing
$m_a,\vec v_a,\vec c_a$ by $m'_a,\vec v'_a,\vec c'_a$ and reversing the overall sign,
where $\vec c'_a$ is the incoming coefficient obtained by the preceding rule. Since the
square bracket in \refb{d_ra} is homogeneous of degree $r$ in the coefficients
$\vec c_a$, the resulting $\vec d'_{r,a}$ is $(-1)^{r+1}$ times the expression obtained
by evaluating \refb{c_a} and \refb{d_ra} directly on the incoming masses and
velocities, with no sign changes. The early-time waveform is then obtained from
\refb{eij_late_time} by replacing $m_a,\vec v_a,\vec c_a,\vec d_{r,a},z$ by
$m'_a,\vec v'_a,\vec c'_a,\vec d'_{r,a},z'$, respectively, where $z':=\ln|w|/w$.
Here $w$ is negative, while the masses $m'_a$ multiplying the waveform keep their
positive sign.

We also note that the $w$-independent part of the waveform can be removed (locally, in
the vicinity of the detector) by a coordinate transformation of the form
$x^\mu \to x^\mu + S^\mu{}_\nu x^\nu/R$, with constant $S^\mu{}_\nu$. However, the
difference between its values in the past and in the future cannot be removed in this
way. It is therefore often convenient to use this freedom to set the $w$-independent
part of the past waveform to zero, in which case the $w$-independent part of the
future waveform becomes
\be\label{eq:memory_frame}
{2 G \alpha^2\over R} \left(\sum_a m_a v_a^i v_a^j - \sum_b m'_b v_b^{\prime i}
v_b^{\prime j}\right)\, .
\ee
This is the gravitational memory.

Specializing the multiparticle coefficients \refb{c_a} and \refb{d_ra} to
two outgoing particles and substituting them into \refb{eij_late_time}
gives a geometric series that can be summed explicitly. With
$\vec v:=\vec v_1-\vec v_2$, $v:=|\vec v|$ and $\gamma:=G(m_1+m_2)/v^3$, the
late-time waveform for $w\to+\infty$ is
\be\label{eq:intro-two}
e^{ij}(w)={2G\alpha^2\over R}\left[
m_1v_1^iv_1^j+m_2v_2^iv_2^j+
{Gm_1m_2\over v^3}\,v^iv^j\,w^{-1}(1+\gamma z)^{-1}\right].
\ee
The factor $w^{-1}(1+\gamma z)^{-1}$ resums all the terms $w^{-k}(\ln w)^{k-1}$ with
$k\ge1$ at leading Newtonian order. For two incoming particles, define
$\vec v':=\vec v'_1-\vec v'_2$, $v':=|\vec v'|$ and
$\gamma':=-G(m'_1+m'_2)/v^{\prime 3}$. The replacement rules stated above give the
early-time waveform, for $w\to-\infty$,
\be\label{eq:intro-two-incoming}
e^{ij}(w)={2G\alpha^2\over R}\left[
m'_1v_1^{\prime i}v_1^{\prime j}+m'_2v_2^{\prime i}v_2^{\prime j}
-{G m'_1m'_2\over v'{}^3}\,v^{\prime i}v^{\prime j}\,
 w^{-1}(1+\gamma'z')^{-1}\right].
\ee

For an arbitrary number of particles, \refb{eij_late_time} reproduces the Newtonian
limits of the constant, $u^{-1}$ and $u^{-2}\ln|u|$ terms fixed by the classical soft
graviton theorem~\cite{1912.06413,2008.04376,2105.08739}. Beyond these known terms,
\refb{eij_late_time}--\refb{d_ra} provide a new determination of the
highest-logarithmic waveform coefficients for three or more asymptotic particles at
leading Newtonian order.

For $2\to2$ scattering, the resummed expressions \refb{eq:intro-two} and
\refb{eq:intro-two-incoming} agree with the leading Newtonian limit of the all-orders
conjecture of~\cite{2407.04128}. This agreement holds at every retained logarithmic
order, but does not establish the conjecture beyond the Newtonian limit. For a specific
three-particle outgoing state, however, our result differs from the conjecture
of~\cite{2407.04128} already in the coefficient of $w^{-3}(\ln w)^2$. The difference
has a nonzero transverse-traceless part, so it changes the radiative waveform and
provides a counterexample to the multiparticle conjecture at leading Newtonian order.

The multiparticle waveform conjecture of~\cite{2407.04128} also fails another
consistency test. Formally replacing one particle by two coincident pieces on its
original trajectory, with its mass and momentum divided in the same proportions,
leaves the total matter energy-momentum tensor unchanged. Mutual interactions between
the pieces are self-interactions of the original particle and are excluded, so the
waveform should be invariant under this replacement. The first three soft
coefficients, which are known, satisfy this requirement, but the conjectured
higher-order coefficients do not. Our Newtonian result \refb{eij_late_time} satisfies
this invariance, providing an additional consistency check on the derived
multiparticle waveform.

After making some ad hoc assumptions, we were also able to obtain an all-orders
expression for the gravitational waveform beyond the Newtonian limit. Since we have not
been able to fully justify all the assumptions that went into its derivation, we do not
give the details of the derivation, and we present this expression as a conjecture
rather than a result. Nevertheless, the conjecture passes many nontrivial consistency
checks. The conjecture and these checks are described in Section~\ref{s5}. We hope to
provide a complete derivation, justifying these assumptions, in future
work~\cite{KKMSS:wip}.

The rest of the paper is organized as follows. Section~\ref{s2} derives the
asymptotic trajectory coefficients and the corresponding early- and late-time
multiparticle waveforms in the Newtonian approximation, together with the first few
terms in the low-frequency expansion of their Fourier transforms.
Section~\ref{S:two_particle} specializes to two particles in each
asymptotic state and resums the retained logarithmic terms. Section~\ref{S:comparisons}
compares these results with the known soft graviton coefficients and with the
all-orders conjecture of~\cite{2407.04128}, establishing two-particle agreement and a
three-particle counterexample. Section~\ref{s5} presents our all-orders conjecture
beyond the Newtonian limit. Appendix~\ref{sa} estimates the corrections
omitted in the Newtonian calculation, and Appendix~\ref{sb} examines the
particle-subdivision test. Appendix~\ref{sc} explains how to extract the
highest-logarithmic coefficients in the presence of lower logarithms.
Appendix~\ref{sd} gives a new explicit expression for $A^{\mu\nu}_{(3)}$, the
coefficient of $u^{-3}(\ln|u|)^2$, obtained from our conjecture.

\newpage
\sectiono{Multiparticle waveform in the Newtonian limit} \label{s2}

We consider a scattering process in which incoming particles with masses $\{m'_a\}$
and asymptotic velocities $\{\alpha\vec v'_a\}$ collide and disperse into outgoing
particles with masses $\{m_a\}$ and asymptotic velocities $\{\alpha\vec v_a\}$. Here
$\alpha\ll1$ is a small parameter, and the masses and rescaled velocities $\vec v_a$,
$\vec v'_a$ are held fixed as $\alpha\to0$. The Newtonian approximation thus corresponds
to small asymptotic velocities. We use natural units with $c=1$ and the Minkowski metric
$\eta_{\mu\nu}=\mbox{diag}(-1,+1,+1,+1)$ in four-dimensional flat spacetime. Apart from
requiring the incoming and outgoing particles to be nonrelativistic, we make no
assumption about the interactions they undergo during the collision. In particular, the
particles may move relativistically at intermediate stages of the scattering; only their
asymptotic velocities need to be small. Our goal will be to compute the gravitational
waveform observed by a distant detector at early and late retarded times $u$. For
definiteness, we shall carry out the analysis at late retarded times; the result at
early retarded times follows from replacement rules given below
\refb{eq:general-dr}.

Let $T^{\mu\nu}(t,\vec x)$ be the total energy-momentum tensor as a function of the
spacetime coordinates $(t,\vec x)$, and let $e^{\mu\nu}(t', \vec x')$ be the
trace-reversed metric fluctuation produced at the spacetime point $(t',\vec x')$ by the
energy-momentum tensor. We define $h_{\mu\nu}$ and $e_{\mu\nu}$ through
\be
g_{\mu\nu} =\eta_{\mu\nu}+2\, h_{\mu\nu}, \qquad e_{\mu\nu}:=h_{\mu\nu}-
{1\over 2}\, \eta_{\mu\nu}\,
\eta^{\rho\sigma} h_{\rho\sigma}, \qquad h_{\mu\nu}=e_{\mu\nu}- {1\over 2}\,
\eta_{\mu\nu}\,
\eta^{\rho\sigma} e_{\rho\sigma}\, ,
\ee
and all indices are raised and lowered with the Minkowski metric $\eta$.
In harmonic gauge, $\p_\mu e^{\mu\nu}=0$, the Einstein equations take the
form $\Box e^{\mu\nu}=-8\pi G\,T^{\mu\nu}$, where $T^{\mu\nu}$ includes the gravitational
contribution. Using the position-space form of the retarded Green's function,
\be
G_{\rm ret}(x', x) = - {1\over 4\pi\,  |\vec x'-\vec x|} \, \delta(t'-t - |\vec x'-\vec x|)\, ,
\ee
we have
\be\label{e10}
e^{\mu\nu}(t',\vec x') = 2\, G\int d^3 x {1\over |\vec x'-\vec x|} T^{\mu\nu}(t(\vec x), \vec x)\, ,
\ee
where $t(\vec x)$, which also depends on the fixed observation point $(t',\vec x')$, is determined by
\be\label{e11}
(t'- t(\vec x))^2 = (\vec x'-\vec x)^2, \qquad t'> t(\vec x)\, .
\ee
We shall take
\be\label{e12}
\vec x' = R\vec n, \qquad \vec n^2 =1, \qquad t' = R + u\, ,
\ee
with $T^{\mu\nu}$ concentrated around $\vec x=0$ at $t=0$. For large $R$, \refb{e11} gives
\be\label{e13}
(R + u - t(\vec x))^2 = (R\vec n-\vec x)^2 = R^2 - 2 R\,\vec n\cdot\vec x + \vec x^2 \, .
\ee
This gives
\be \label{e14}
t(\vec x) = u + \vec n\cdot\vec x + O(1/R)\, .
\ee
Since we take the large-$R$ limit and keep only the leading power of $1/R$,
we can ignore the $O(1/R)$ term and write
\be\label{e15}
e^{\mu\nu}(t',\vec x') = {2\, G\over R} \int d^3 x  T^{\mu\nu}(u + \vec n\cdot\vec x,
\vec x)\, .
\ee

We claim that in the Newtonian limit, where the asymptotic velocities are small, we can ignore
the $\vec n\cdot\vec x$ term. We therefore write
\be\label{e16}
e^{\mu\nu}(t',\vec x') = {2\, G\over R}
\int d^3 x\,  T^{\mu\nu}(u,
\vec x) + \EE_1^{\mu\nu}\, ,
\ee
where
$\EE_1^{\mu\nu}$ is the error term
\be  \label{e1.9}
\EE_1^{\mu\nu} = {2\, G\over R}
\int d^3 x  \{ T^{\mu\nu}(u + \vec n\cdot\vec x,
\vec x) - T^{\mu\nu}(u,
\vec x)\}\, .
\ee
We shall show in Appendix~\ref{sa.1} that, in the desired approximation,
$\EE_1^{\mu\nu}$ can be ignored. For now, we shall proceed by dropping this term. At
this order, the first term on the right-hand side of \refb{e16} has no explicit
dependence on $\vec n$. The observed transverse-traceless waveform nevertheless depends
on the detector direction through the transverse-traceless projection.

The total energy-momentum tensor is conserved,
\be
\p_\mu T^{\mu\nu}=0\, .
\ee
In particular, $\p_tT^{00}=-\p_iT^{i0}$ and $\p_tT^{i0}=-\p_jT^{ji}$, with $i,j=1,2,3$
spatial indices. Differentiating the first relation with respect to $t$ and using the
second yields
\be
\p_t^2 T^{00}=\p_i\p_jT^{ij}\, .
\ee
This gives
\be
\p_t^2 \int d^3 x \, x^k x^l T^{00}(t,\vec x)
= \int d^3 x \, x^k x^l \p_i \p_j T^{ij}(t,\vec x)
= 2\int d^3 x \, T^{kl}(t,\vec x)\, ,
\ee
where in the last step we have integrated by parts twice.\footnote{With $T^{00}$ the total energy density, the integral
$\int d^3x\,x^kx^lT^{00}$ diverges linearly at large $r$, since the gravitational
contribution falls off only as $O(r^{-4})$. The divergent part comes from the monopole
part of the far field and is proportional to $GM^2\delta^{kl}$. Since the total mass $M$
is conserved, the second time derivative annihilates it, so the relation above holds with
the derivatives taken before the large-$r$ cutoff is removed.\label{fn:quad-div}} The boundary terms at infinity
vanish, since $\p_j T^{ij}(t,\vec x)=O(r^{-5})$ and $T^{ij}(t,\vec x)=O(r^{-4})$ for large
$r=|\vec x|$. Since the integration region includes the particle positions, where the
integrand has delta-function contributions, there are no boundary terms from the
particle positions either. Combining this with \refb{e16}, with $\EE_1^{\mu\nu}$
dropped, we obtain the quadrupole formula~\cite{0501041,1310.1528}
\be \label{e1.13}
e^{ij}(t',\vec x') =
{G\over  R} {d^2\over du^2}
\int d^3 x\,  x^i \, x^j \,T^{00}(u,
\vec x) \, .
\ee
Our strategy is to use this relation to calculate $e^{ij}$, and then to find
$e^{0\nu}$ using the relation\footnote{At radiative order, the harmonic gauge condition
reduces to $n_\mu \frac{d}{du}e^{\mu\nu}=0$. Since the $u$-independent piece can be
removed by a coordinate transformation, as discussed around \eqref{eq:memory_frame},
for the time-dependent part of the waveform it reduces to
\eqref{eq:gauge_condition}. \label{fo1}}
\be\label{eq:gauge_condition}
n_\mu e^{\mu\nu}=0, \qquad n:= (1, \vec n)\, .
\ee

Since the right-hand side of \refb{e1.13} depends on the observation point
$(t',\vec x')=(R+u,R\vec n)$ only through $u$ and the overall factor $1/R$, from now on
we write the waveform as $e^{ij}(u)$, leaving the dependence on $R$ implicit.
Let $\vec X_a(t)$ be the trajectory of the $a$'th particle. Then \refb{e1.13} becomes
\be\label{eq:quad-particle}
e^{ij}(u)= {G\over  R}  {d^2 \over du^2} \sum_a m_a X_a^{i}(u) X_a^{j}(u)
+ \EE_2^{ij}\, ,
\ee
where the error term
\be \label{e1.17}
\EE_2^{ij}= {G\over R} {d^2\over du^2}\left[
 \int d^3 x x^i x^j T^{00}(u,\vec x) -  \sum_a m_a X_a^{i}(u) X_a^{j}(u) \right]
\ee
comes from the kinetic energies of the particles and the gravitational energy
density. We shall show in Appendix~\ref{sa.2} that, in the desired approximation,
$\EE_{2}^{ij}$ can be ignored. For now, we proceed by dropping this term and write
\be \label{e1.16}
e^{ij}(u)  = {G\over  R} \sum_a m_a \left[ 2 {d X_a^i(u) \over du} {d X_a^j(u) \over du}
+ X_a^{i}(u) {d^2 X_a^j(u) \over du^2} + X_a^{j}(u) {d^2 X_a^i(u) \over du^2}
\right]\, .
\ee
The first term in \refb{e1.16} comes from the matter contribution to $T^{ij}$, while
the last two terms come from the gravitational contribution to $T^{ij}$, expressed in
terms of the matter trajectories via the conservation of $T^{\mu\nu}$.

To make the small-$\alpha$ expansion manifest,\footnote{The power of $\alpha$ in \refb{e1.18} reflects the scaling
symmetry of Newtonian gravity, $\vec X\to\lambda^2\vec X$, $t\to\lambda^3 t$: with velocities of
order $\alpha$, the natural length and time scales are $\alpha^{-2}$ and
$\alpha^{-3}$.} we define the rescaled retarded time
$w$
and the variable $z$ by
\be\label{e1.18}
w := \alpha^3 u, \qquad z := {\ln w\over w}\, .
\ee
Regarding $\vec X_a$ as a function of $w$, \refb{e1.16} becomes
\be \label{e1.16a}
e^{ij}(w)  ={G\alpha^6
\over R} \sum_a m_a \left[ 2\, {d X_a^i \over dw} {d X_a^j \over dw}
+ X_a^{i} {d^2 X_a^j \over dw^2} + X_a^{j} {d^2 X_a^i \over dw^2}
\right]\, .
\ee

We shall work in the following approximations:
\begin{enumerate}
\item We take $R$ to be large and keep only terms of order $1/R$, as in \refb{e15}.
\item \label{item:formal-limit} We formally work in the limit of large $w$ at fixed $z$.
This means that we allow higher powers of $1/w$ as long as each additional
power of $1/w$ is accompanied by a power of $\ln w$.
In this counting, we keep terms of order $w$ and order unity in $\vec X_a$, up
to order $w^{-1}$ in $d\vec X_a/dw$, up to order $w^{-2}$ in $d^2\vec X_a/dw^2$, and up
to order $w^{-1}$ in $e^{ij}$. In the waveform, this selects the constant term and the
terms $w^{-1}z^{k-1}=w^{-k}(\ln w)^{k-1}$, $k\ge1$. It is a formal prescription rather than a
truncation by magnitude: for example, $w^{-3}(\ln w)^2$ is retained while $w^{-2}$ is
omitted. Lower logarithms are left undetermined. Appendix~\ref{sc} explains how to
extract the retained coefficients in the presence of these terms.
\item In the small-$\alpha$ expansion, carried out at fixed $w$ and $z$, we
retain the leading $O(\alpha^{-2})$ terms in $\vec X_a$, $d\vec X_a/dw$ and
$d^2\vec X_a/dw^2$, keeping in each only the orders in $w$ specified in item~\ref{item:formal-limit}.
Terms of order $\alpha^{-1}$ and higher
in these trajectory quantities are discarded. Consequently, we retain the
$O(\alpha^2)$ spatial waveform and neglect corrections of order $\alpha^3$
and higher.
\end{enumerate}
We shall show in Appendices~\ref{sa.1} and~\ref{sa.2} that, in this approximation,
$\EE_1^{ij}$ and $\EE_2^{ij}$ do not contribute at this order. Hence we can work
with \refb{e1.16a}.

We begin with the ansatz that the late-time trajectory of the $a$'th particle takes the
form
\be \label{e1.20}
\vec X_a(u) =\alpha^{-2}\, \left[\left\{ \vec v_a \, w +  \vec c_a\, w\, z +
\vec b_a + \sum_{r\ge 1}
\vec d_{r,a} z^r + \OO(w^{-1}) \right\} + \OO(\alpha)\right]
\ee
with
\be
|\vec v_a|\sim 1, \qquad |\vec c_a|\sim 1, \qquad |\vec b_a|\sim 1, \qquad |\vec d_{r,a}|\sim 1\, .
\ee
We shall see shortly that the equations of motion admit a solution of this form.
Differentiating, and remembering that $z$ depends on $w$ through \refb{e1.18}, we obtain
\be \label{e2.23zz}
{d\vec X_a\over dw} = \alpha^{-2}\, \left[\vec v_a + \vec c_a\, w^{-1}
- w^{-1} \sum_{r\ge 1}
\vec d_{r,a}  r z^r + \OO(w^{-2})\right] + \OO(\alpha^{-1})\, ,
\ee
\be\label{e1.23}
{d^2\vec X_a\over dw^2} = \alpha^{-2}\, \left[ - \vec c_a\, w^{-2}
+ w^{-2} \sum_{r\ge 1}
\vec d_{r,a}  r(r+1) z^r +\OO(w^{-3})\right] + \OO(\alpha^{-1})\, .
\ee
We can substitute these into \refb{e1.16a} to find $e^{ij}$ to the desired order. Note
that, to get a contribution of order $w^0$ or $w^{-1}$, we need to pick the $\vec v_a$
term from $d\vec X_a/dw$ in at least one of the two factors in the first term, and the
$\vec v_a w + \vec c_a w z$ terms from $\vec X_a$ in the last two terms. Other terms are
of order $w^{-2}$ and can be ignored in our approximation. This gives
\ben \label{e1.25}
e^{ij}(w) &=& {G\alpha^2\over  R} \sum_{a} m_a  \Bigg[ 2 v_a^i v_a^j
+ 2 w^{-1} \, v_a^i \Bigg(c_a^j - \sum_{r\ge 1} d^j_{r,a} r z^r\Bigg)
+ 2 w^{-1} \, v_a^j \Bigg(c_a^i - \sum_{r\ge 1} d^i_{r,a} r z^r\Bigg) \nonumber \\
&+& w^{-1} \Bigg(v_a^i + c_a^i \, z\Bigg) \Bigg(  -c_a^j + \sum_{r\ge 1} d_{r,a}^j r(r+1)
z^r\Bigg) \nonumber \\
&+& w^{-1} \Bigg(v_a^j + c_a^j \, z\Bigg) \Bigg(  -c_a^i + \sum_{r\ge 1} d_{r,a}^i r(r+1)
z^r\Bigg)\Bigg]\, .
\een
This is the result \refb{eij_late_time} quoted in the introduction. It is of order
$\alpha^2$ as long as the coefficients $\vec c_a$ and $\vec d_{r,a}$ are of order unity.

We now determine the coefficients $\vec c_a$ and $\vec d_{r,a}$. For this, we use the
equations of motion
\be \label{e1.25xxy}
m_a {d^2 \vec X_a\over dw^2} = \alpha^{-6} m_a {d^2 \vec X_a\over du^2}
= -\alpha^{-6}  G\,  \sum_{b\ne a}  m_a m_b {\vec X_a - \vec X_b\over |\vec X_a-
\vec X_b|^3} + \EE_3\, ,
\ee
where $\EE_3$ is an error term that accounts for forces other than the Newtonian
gravitational force. We show in Appendix~\ref{sa.3} that $\EE_3$ does not contribute
at the desired order. Dropping it, dividing by $m_a$ and using \refb{e1.20} and
\refb{e1.23}, we get
\ben\label{e1.27}
&&\alpha^{-2}\, \left[ - \vec c_a\, w^{-2}
+ w^{-2} \sum_{r\ge 1}
\vec d_{r,a}  r(r+1) z^r +\OO(w^{-3})\right]
\nonumber \\
&=& -\alpha^{-2}\, G\,  \sum_{b\ne a} m_b {(\vec v_a-\vec v_b)\, w + (\vec c_a-
\vec c_b)\,  w\, z + (\vec b_a-\vec b_b)
+ \sum_{r\ge 1} (\vec d_{r,a}-\vec d_{r,b}) z^r \over
|(\vec v_a-\vec v_b)\, w + (\vec c_a-\vec c_b)\,  w\, z + (\vec b_a-\vec b_b)
+ \sum_{r\ge 1} (\vec d_{r,a}-\vec d_{r,b}) z^r|^3
} \nonumber \\
&=& -\alpha^{-2}\,
G\,   w^{-2} \sum_{b\ne a} m_b{(\vec v_a-\vec v_b) + (\vec c_a-\vec c_b)\,  z  \over
|(\vec v_a-\vec v_b) + (\vec c_a-\vec c_b)\,  z |^3
} + \OO(w^{-3})\, .
\een
The factors of $\alpha^{-2} w^{-2}$ cancel on both sides, and the vectors $\vec b_a$
drop out at this order. Expanding the right-hand side in powers of $z$ and matching
coefficients with the left-hand side, we obtain
\begin{align}
\vec c_a &= G\, \sum_{b\ne a}  m_b\, {\vec v_a -\vec v_b\over |\vec v_a-\vec v_b|^3}\, ,
\label{e1.28y}\\*[2mm]
\vec d_{r,a} &= -{G\over 2^r r(r+1)}\sum_{b\ne a}
{m_b\over |\vec v_a-\vec v_b|^3}
\Bigg[ (\vec v_a-\vec v_b)
\sum_{\ell=[(r+1)/2]}^{r}
{(-1)^\ell(2\ell+1)!\over \ell!\,(2\ell-r)!\,(r-\ell)!}
\notag\\*
&\hskip 1.2in\times
{\bigl[(\vec v_a-\vec v_b)\cdot(\vec c_a-\vec c_b)\bigr]^{2\ell-r}
|\vec c_a-\vec c_b|^{2(r-\ell)}\over |\vec v_a-\vec v_b|^{2\ell}}
\notag\\*
&\hskip 0.9in+2(\vec c_a-\vec c_b)
\sum_{\ell=[r/2]}^{r-1}
{(-1)^\ell(2\ell+1)!\over \ell!\,(2\ell+1-r)!\,(r-\ell-1)!}
\notag\\*
&\hskip 1.2in\times
{\bigl[(\vec v_a-\vec v_b)\cdot(\vec c_a-\vec c_b)\bigr]^{2\ell+1-r}
|\vec c_a-\vec c_b|^{2(r-\ell-1)}\over |\vec v_a-\vec v_b|^{2\ell}}
\Bigg],\qquad r\ge1\, ,
\label{eq:general-dr}
\end{align}
which are the expressions quoted in \refb{c_a} and \refb{d_ra}.

A similar formula holds for the waveform at large negative $w$, with $m_a$,
$\vec v_a$ and $z$ replaced by $m'_a$, $\vec v'_a$ and $z':=\ln|w|/w$. There is, however,
one important difference in the computation of the trajectories. In going from the
second to the third line of \refb{e1.27}, we pulled a factor of $w$ out of the numerator
and $|w|^3$ out of the denominator, and used $w/|w|^3=w^{-2}$. For negative $w$, we
instead have $w/|w|^3=-w^{-2}$. This sign can be compensated by reversing the signs of
all the masses when computing the trajectories. Thus, for the incoming particles, we
replace $m_a$ by $-m'_a$ rather than by $m'_a$ in \refb{e1.28y} and
\refb{eq:general-dr}; for example, $\vec c'_a$ acquires an extra minus sign relative to
\refb{e1.28y}. The masses $m'_a$ multiplying the waveform in \refb{e1.25}, on the other
hand, remain positive.

Equations~\refb{e1.25}, \refb{e1.28y} and \refb{eq:general-dr},
together with the replacement rules just described, are the main results of this paper
in the Newtonian limit. They determine the memory term and all the highest-logarithmic
terms $w^{-k}(\ln|w|)^{k-1}$, $k\ge1$, of the late- and early-time waveforms for an
arbitrary number of particles, in terms of their masses and asymptotic velocities
alone.

For comparison with known results, we now extract the first three terms of
\refb{e1.25}: the constant term and the terms proportional to $w^{-1}$ and
$w^{-2}\ln w=w^{-1}z$. Keeping only these terms, we have
\be \label{e1.25x}
e^{ij}(w) = {G\alpha^2\over  R} \sum_{a} m_a  \Bigg[ 2\, v_a^i \, v_a^j
+ w^{-1} \, (v_a^i \, c_a^j + v_a^j \, c_a^i) - 2 \, w^{-1}\, z \, c_a^i \, c_a^j \Bigg]\, .
\ee
The contributions from $\vec d_{1,a}$ cancel at this order, so \refb{e1.25x}
depends only on $\vec v_a$ and $\vec c_a$.

Since some of the relativistic results are stated in frequency space, we also give the
low-frequency behavior of the Fourier transform of the waveform. We define the Fourier
transform of $e^{ij}(w)$ by\footnote{Due to the rescaling $w=\alpha^3 u$, $\nu$ is
related to the frequency $\omega$ conjugate to $u$ by $\nu=\omega/\alpha^3$. The factor
$\alpha^{-3}$ ensures that $\tilde e^{ij}$ equals the Fourier transform with respect to
$u$, $\int du\, e^{i\omega u} e^{ij}$, so that it can be compared directly with the
relativistic results.}
\be
\tilde e^{ij} (\nu) = \alpha^{-3}\int_{-\infty}^\infty \, dw \, e^{i\nu w} \, e^{ij}(w)\, ,
\qquad e^{ij}(w) = \alpha^3
\int_{-\infty}^\infty \, {d\nu\over 2\pi} \, e^{-i\nu w} \,  \tilde e^{ij} (\nu)\, .
\ee
Using the relations \cite{2008.04376}
\ben \label{e1.30}
&& \int_{-\infty}^\infty {d\nu\over 2\pi}\, e^{-iw\nu}\, \nu^{n-1}\,
(\ln(\nu+i\eps))^{n-r} (\ln(\nu-i\eps))^r \nonumber \\
&\simeq& \begin{cases} - (n-r) \, (n-1)! \, i^{n-1} \, {(\ln|w|)^{n-1} /  w^n} & \hbox{for
$w\to\infty$}\\  r \, (n-1)! \, i^{n-1} \, {(\ln|w|)^{n-1} /  w^n} & \hbox{for
$w\to-\infty$}\end{cases} \, ,
\een
for $n\ge 1$ and $0\le r\le n$, and
\ben \label{e1.31}
&& \int_{-\infty}^\infty {d\nu\over 2\pi}\, e^{-i w\nu}\, (\nu+i\eps)^{-1}
=\begin{cases} -i  & \hbox{for
$w\to\infty$}\\ 0 & \hbox{for
$w\to-\infty$}\end{cases} \nonumber \\
&& \int_{-\infty}^\infty {d\nu\over 2\pi}\, e^{-i w\nu}\, (\nu-i\eps)^{-1}
=\begin{cases}  0 & \hbox{for
$w\to \infty$}\\
i & \hbox{for
$w\to-\infty$}
\end{cases}\, ,
\een
we find that the late-time behavior \refb{e1.25x}, together with its early-time
counterpart, determines the leading nonanalytic terms of $\tilde e^{ij}(\nu)$ at small
$\nu$:
\ben \label{efourier}
\tilde e^{ij}(\nu) &\simeq& {2\, G\over  R\, \alpha} \Bigg[ \sum_a m_a \, v_a^i \, v_a^j \,
 {i\over \nu+i\eps} -
\sum_a m_a' v_a^{\prime i} v_a^{\prime j}   {i\over \nu-i\eps}
\Bigg] \nonumber \\
&-& {G\over  R\, \alpha} \Bigg[ \sum_a m_a \, (v_a^i \, c_a^j +v_a^j c_a^i)
\,
 \ln(\nu+i\eps) - \sum_a m_a' \, (v_a^{\prime i} \, c_a^{\prime j} +v_a^{\prime j}
 c_a^{\prime i})
\,
 \ln(\nu- i\eps)
\Bigg] \nonumber \\
&-& { i\, G\over  R\, \alpha} \Bigg[ \sum_a m_a \, c_a^i \, c_a^j \,
\, \nu \,
 (\ln(\nu+i\eps))^2  - \sum_a m_a' \, c_a^{\prime i} \, c_a^{\prime j} \,
\, \nu \,
 (\ln(\nu-i\eps))^2 \Bigg]\, .
 \een
Equation~\refb{efourier} gives the frequency-space counterparts of the constant,
$w^{-1}$ and $w^{-2}\ln|w|$ terms at leading Newtonian order, in the sense of the
asymptotic expansion \refb{eq:waveform-log-expansion} and the coefficient extraction
\refb{eq:waveform-log-projector}. Here and below, asymptotic equalities retain only the
leading poles and logarithms; lower logarithms and terms regular in the frequency are
left unspecified.

\sectiono{Two particles in the initial and final states}\label{S:two_particle}

In this section, we specialize the results of Section~\ref{s2} to the case
of two particles in each asymptotic state, where the highest-logarithmic terms of the
waveform can be summed in closed form.
For two outgoing particles, the matching condition \refb{e1.27} can be solved with the
ansatz
\be\label{eq:two-body-ansatz}
\vec v := \vec v_1 - \vec v_2, \qquad \vec c_a = c_a \, \vec v,  \qquad \vec d_{r,a} = d_{r,a}\,
\vec v, \qquad \hbox{for $a=1,2$}\, .
\ee
Here $c_a$ and $d_{r,a}$ are scalars, not to be confused with the
components $c_a^i$ and $d_{r,a}^i$ of the vectors $\vec c_a$ and $\vec d_{r,a}$.
Substituting this into \refb{e1.27}, with $v:=|\vec v|$, we get
\ben\label{e2.1}
&& -   c_1
+ \sum_{r\ge 1}
  d_{r,1} \,  r(r+1) \, z^r
= -
{G\, m_2\over v^3} {1 + (  c_1-  c_2)\,  z  \over
|1 + (  c_1-  c_2)\,  z |^3 }  \nonumber \\
&& -   c_2
+ \sum_{r\ge 1}
  d_{r,2} \,  r(r+1) \, z^r
=
{G\, m_1\over v^3} {1 + (  c_1-  c_2)\,  z  \over
|1 + (  c_1-  c_2)\,  z |^3 } \, .
\een
This gives
\be \label{e2.3z}
c_1= {m_2 \over m_1+m_2} \gamma, \qquad
c_2 = -  {m_1 \over m_1+m_2} \gamma, \qquad \gamma :=
{G\, (m_1+m_2)\over v^3}\, ,
\ee
\be\label{eq:two-body-dr}
d_{r,1} = -(-1)^r \, {1\over r} \, {m_2 \over m_1+m_2}  \, \gamma^{r+1}
\, ,
\qquad
d_{r,2} = (-1)^r \, {1\over r} \,   {m_1 \over m_1+m_2}  \,  \gamma^{r+1}
 \, .
\ee
Equations \refb{e2.3z} and \refb{eq:two-body-dr} also follow by
specializing the general multiparticle expressions \refb{e1.28y} and
\refb{eq:general-dr}, respectively, to two particles and substituting
\refb{eq:two-body-ansatz}.

Substituting \refb{e2.3z} and \refb{eq:two-body-dr} into \refb{e1.25}, we get
\be \label{e1.25b}
e^{ij}(w) = {2\, G\alpha^2\over  R} \Bigg[ m_1 \, v_1^i\, v_1^j + m_2 \, v_2^i
\, v_2^j
+ {m_1m_2\over m_1+m_2} \,
v^i \, v^j \,  w^{-1} \, \gamma \, \sum_{k=0}^\infty (-1)^k \gamma^k z^k
\Bigg]\,.
\ee
Summing the geometric series, we obtain the closed-form expression
\be \label{e1.25a}
e^{ij}(w) = {2\, G\alpha^2\over  R} \Bigg[ m_1 \, v_1^i\, v_1^j + m_2 \, v_2^i
\, v_2^j
+ {G\, m_1m_2\over v^3} \,
v^i \, v^j \,  w^{-1} \,  (1+\gamma\, z)^{-1}
\Bigg]\, ,
\ee
with the understanding that it is to be expanded formally in powers of $z$. The
geometric series in \refb{e1.25b} sums to \refb{e1.25a} for $|\gamma z|<1$;
successive terms are parametrically smaller when
$|\gamma z|=G(m_1+m_2)|\ln w|/(v^3w)\ll1$.

Equation~\refb{e1.25b} gives the waveform at large positive $w$. If the incoming state
also consists of two particles, with masses $m'_1$ and $m'_2$ and asymptotic velocities
$\alpha\vec v'_1$ and $\alpha\vec v'_2$, then, by the replacement rules discussed below
\refb{eq:general-dr}, the waveform at large negative $w$ is
\be \label{e1.25past}
e^{ij}(w) = {2\, G\alpha^2\over  R} \Bigg[ m'_1 \, v_1^{\prime i}\, v_1^{\prime j} + m'_2 \,
v_2^{\prime i}
\, v_2^{\prime j}
+ {m'_1m'_2\over m'_1+m'_2} \,
v^{\prime i} \, v^{\prime j} \,  w^{-1} \, \gamma' \, \sum_{k=0}^\infty (-1)^k (\gamma')^k
z^{\prime k}
\Bigg]\, ,
\ee
where $\vec v':=\vec v'_1-\vec v'_2$, $v':=|\vec v'|$ and
\be
\gamma' := -{ G(m_1'+m_2')\over v^{\prime 3}}\, , \qquad z' := {\ln|w|\over w}\, .
\ee
The minus sign in $\gamma'$ is as discussed below \refb{eq:general-dr}.
The resummed forms of \refb{e1.25b} and \refb{e1.25past} are the
expressions \refb{eq:intro-two} and \refb{eq:intro-two-incoming} quoted in the
introduction.

As in Section~\ref{s2}, the late- and early-time behavior \refb{e1.25b} and
\refb{e1.25past}, together with \refb{e1.30} and \refb{e1.31}, determines the leading
nonanalytic terms of the Fourier transform at small $\nu$:
\ben \label{e2.9xx}
\tilde e^{ij}(\nu) &\simeq&  {2\, G\over  R\, \alpha} \Bigg[ (m_1 \, v_1^i\, v_1^j + m_2 \, v_2^i
\, v_2^j ) {i\over \nu+i\eps} -
(m'_1 \, v_1^{\prime i}\, v_1^{\prime j} + m_2' \,
v_2^{\prime i}
\, v_2^{\prime j})  {i\over \nu-i\eps}
\Bigg] \nonumber \\
&-& {2\, G\over  R\, \alpha}  v^i \, v^j \,
\sum_{k=1}^\infty {1\over k!}  \, i^{k-1} \, {m_1m_2\over m_1+m_2} \left({G\, (m_1+m_2)\over v^3}\right)^{k}\, \nu^{k-1}\, (\ln(\nu+i\eps))^{k}
\nonumber \\
&+& {2\, G\over  R\, \alpha}  v^{\prime i} \, v^{\prime j} \,
\sum_{k=1}^\infty {1\over k!}  \, i^{k-1} \, {m'_1m'_2\over m'_1+m'_2} \left(-{G\, (m_1'+m_2')\over v^{\prime 3}}\right)^{k}
\, \nu^{k-1}\, (\ln(\nu-i\eps))^{k}\, .
\een
In the next section, we compare these results, together with the
multiparticle results of Section~\ref{s2}, with known soft graviton results and with
the all-orders conjecture of~\cite{2407.04128}.

\sectiono{Comparison with known and conjectured results}\label{S:comparisons}

In this section, we compare our results with the all-orders conjecture
of~\cite{2407.04128} for the scattering of incoming particles with momenta $\{p_a'\}$
into outgoing particles with momenta $\{p_a\}$, with $p_a = (E_a, \vec p_a)$ and
$p'_a=(E'_a, \vec p'_a)$:\footnote{The $1/\omega$ term in the expansion of \refb{e5.4},
proportional to $A_{(0)}^{\mu\nu}-B_{(0)}^{\mu\nu}$, is ambiguous, since we need an
$i\eps$ prescription to determine, via \refb{e1.31}, whether this term is nonzero in the
future or in the past. However, as discussed around \refb{eq:memory_frame}, the only
observable quantity is the difference between the constant parts of the waveform in the
future and in the past.
For definiteness, we take $1/\omega$ to be $1/(\omega+i\eps)$ in the
$A^{\mu\nu}_{(0)}$ term and $1/(\omega-i\eps)$ in the $B^{\mu\nu}_{(0)}$ term, so that the
outgoing and incoming constants appear in the future and in the past, respectively, as
in~\refb{efourier}.}
\ben\label{e5.4}
\tilde e^{\mu\nu}(\omega) &=& -
\exp\left\{-{2\, i\, G } \,\omega\, \ln\, (\omega+i\eps)\,
 {\sum_a}p_a\cdot n \right\}\,  \sum_{k\ge 0}{1\over k!}
  A^{\mu\nu}_{(k)} \, (-i\, \omega)^{k-1} (\ln\,
(\omega+i\eps))^k
\nonumber \\ &&
 -  \exp\left\{-{2\, i\, G } \,\omega\, \ln\, (\omega+i\eps)\,
 {\sum_a}p'_a\cdot n \right\}\,  \sum_{k\ge 0}{1\over k!}
  B^{\mu\nu}_{(k)} \, (i\, \omega)^{k-1} (\ln\,
(\omega-i\eps))^k\, ,
\een
where
\be \label{eamunudef}
A^{\mu\nu}_{(0)} ={2\, G\over R} \,{\sum_{a}}
p_a^\mu \, p_a^\nu\,
{1\over n\cdot p_a}, \qquad B^{\mu\nu}_{(0)} ={2\, G\over R} \, \sum_{a}
p_a^{\prime\mu} \, p_a^{\prime\nu}\,
{1\over n\cdot p_a'} \, ,
\ee
\be \label{ememag}
A^{\mu\nu}_{(1)} =  {(2\, G)^{2}\over  R} \, {\sum_{a,b\atop b\ne a}}
{\tau_{ab}
  \over p_a\cdot n} n_\rho \, p^\mu_{[a} p^\rho_{b]}\,
 p^\nu_{a} \, ,
\qquad
B^{\mu\nu}_{(1)} = {(2\, G)^{2}\over  R} \, \sum_{a,b\atop b\ne a}
{\tau'_{ab}
  \over p'_a\cdot n} n_\rho \, p^{\prime\mu}_{[a} p^{\prime\rho}_{b]}\,
 p^{\prime\nu}_{a} \, ,
\ee
\ben\label{eguess}
A^{\mu\nu}_{(k)} &=& { (2\, G)^{k+1}\over  R}
{\sum_{a,a_1,a_2,\cdots, a_k\atop a_1,\cdots, a_k\ne a} }
{\tau_{aa_1}
 \tau_{aa_2} \cdots
\tau_{aa_k} \over p_a\cdot n} n_\rho \, p^\mu_{[a} p^\rho_{a_1]}\,
n_\sigma p^\nu_{[a} p^\sigma_{a_2]} \nonumber \\* && \hskip 1in \times \
n_{\alpha_3} (p^{\alpha_3}_a+p^{\alpha_3}_{a_3}) \cdots
n_{\alpha_k} (p^{\alpha_k}_a+p^{\alpha_k}_{a_k})\, ,
\een
\ben\label{eguesstwo}
B^{\mu\nu}_{(k)} &=&{(2\, G)^{k+1}\over R}
\sum_{a,a_1,a_2,\cdots, a_k\atop a_1,\cdots, a_k\ne a} {\tau^{\prime}_{aa_1}
 \tau^{\prime}_{aa_2} \cdots
\tau^{\prime}_{aa_k} \over p'_a\cdot n} n_\rho \, p^{\prime\mu}_{[a} p^{\prime\rho}_{a_1]}
\,
n_\sigma \, p^{\prime\nu}_{[a} p^{\prime\sigma}_{a_2]} \nonumber \\* && \hskip 1in \times \
n_{\alpha_3} (p^{\prime\alpha_3}_a+p^{\prime\alpha_3}_{a_3}) \cdots
n_{\alpha_k} (p^{\prime\alpha_k}_a+p^{\prime\alpha_k}_{a_k})\, .
\een
\ben \label{eq:tau-exact}
\tau_{ab} &:=& {1\over 2}\, p_a\cdot p_b\ {2(p_a\cdot p_b)^2 - 3 m_a^2 m_b^2 \over
\{(p_a\cdot p_b)^2 - m_a^2
m_b^2 \}^{3/2}}\, ,
\nonumber \\
\tau'_{ab} &:=& {1\over 2}\,  p'_a\cdot p'_b\ {2(p'_a\cdot p'_b)^2 - {3} m_a^{\prime 2} m_b^{\prime 2}
\over  \{(p'_a\cdot p'_b)^2 - m_a^{\prime 2}
m_b^{\prime 2} \}^{3/2}}  \, ,
\een
\be\label{eq:antisym}
P_{[a}Q_{b]} :=  (P_aQ_b-P_bQ_a)\, .
\ee
Equations~\refb{eamunudef}--\refb{eq:antisym} constitute the all-orders
conjecture of~\cite{2407.04128} for the coefficients $A^{\mu\nu}_{(k)}$ and
$B^{\mu\nu}_{(k)}$ in \refb{e5.4}.
For $0\le k\le 2$, the expressions for $A_{(k)}^{\mu\nu}$ and $B_{(k)}^{\mu\nu}$ given
above were derived earlier in~\cite{1912.06413,2008.04376,2105.08739}, so these terms
provide a necessary consistency check on our results.

In the nonrelativistic limit, we have
\be
\vec p_a \simeq \alpha\, m_a\, \vec v_a, \quad \vec p'_a \simeq \alpha\, m'_a\, \vec v'_a,
\quad p^0_a \simeq m_a, \quad p^{\prime 0}_a \simeq m'_a, \quad
n\cdot p_a \simeq -m_a, \quad n\cdot p'_a \simeq -m'_a\, ,
\ee
\be
(p_a\cdot p_b)^2 - p_a^2 p_b^2 \simeq  m_a^2 m_b^2 \alpha^2 (\vec v_a-\vec v_b)^2,
\qquad (p_a'\cdot p_b')^2 - p_a^{\prime 2} p_b^{\prime 2} \simeq  m_a^{\prime 2}
 m_b^{\prime 2} \alpha^2 (\vec v'_a-\vec v'_b)^2,
\ee
\be \label{etauab}
\tau_{ab} \simeq {1\over 2\, \alpha^3 |\vec v_a -\vec v_b|^3}, \qquad
\tau'_{ab} \simeq {1\over 2\, \alpha^3 |\vec v'_a -\vec v'_b|^3}\, ,
\ee
\ben \label{etauabid}
\sum_{b\ne a}  \tau_{ab} n_\rho p_{[a}^i p_{b]}^\rho &\simeq & -{1\over 2\alpha^2} \sum_{b\ne a}
{1\over |\vec v_a-\vec v_b|^3} m_a m_b (v_a^i-v_b^i)  = -{m_a\over 2\, G\, \alpha^2} c_a^i,
\nonumber \\
 \sum_{b\ne a}  \tau'_{ab} n_\rho p_{[a}^{\prime i} p_{b]}^{\prime \rho} &\simeq&
{m'_a\over 2\, G\, \alpha^2} c_a^{\prime i}\, ,
\een
where we have used \refb{e1.28y} in the last step.
We also substitute
\be \label{e3.10}
\omega\pm i\eps = \alpha^3\, (\nu\pm i\eps),
\qquad \ln(\omega\pm i\eps) \simeq \ln(\nu\pm i\eps)\, .
\ee
We have dropped the $3\ln\alpha$ term from $\ln(\omega\pm i\eps)$, since it only
generates lower powers of the logarithm, which we do not retain. Since
$\omega=\alpha^3\nu$, the exponents in \refb{e5.4} are suppressed by $\alpha^3$, and the
exponential factors can be set to 1. In this limit, \refb{eamunudef}--\refb{eguesstwo}
reduce to
\be\label{e3.13z}
A_{(0)}^{ij} = - {2 G \alpha^2 \over R} \sum_a m_a v_a^i v_a^j, \qquad
B_{(0)}^{ij} = - {2 G \alpha^2 \over R} \sum_a m'_a v_a^{\prime i} v_a^{\prime j},
\ee
\be\label{e3.14z}
A_{(1)}^{ij} = { 2\, G\over  R\, \alpha} \, \sum_{a} \, m_a \, c_a^i
\, v_a^j ,\qquad
B_{(1)}^{ij} = -{ 2\, G\over  R\, \alpha} \, \sum_{a} m_a'
c_a^{\prime i}  \, v_a^{\prime j} \, ,
\ee
\be \label{e3.15z}
A_{(2)}^{ij} = -{ 2\, G\over  R\, \alpha^4} \, \sum_{a} \, m_a \, c_a^i
\, c_a^j ,\qquad
B_{(2)}^{ij} = -{ 2\, G\over  R\, \alpha^4} \, \sum_{a} m_a'
c_a^{\prime i}  \, c_a^{\prime j} \, ,
\ee
\ben \label{e3.16}
&& A_{(k)}^{ij} = { (2\, G)^{k+1}\over  R} \, \alpha^{-3k+2}\, 2^{-k} (-1)^{k+1}
\nonumber \\
&& {\sum_{a,a_1,a_2,\cdots, a_k\atop a_1,\cdots, a_k\ne a} } \prod_{\ell=1}^k |\vec v_a
-\vec v_{a_\ell}|^{-3} \, m_a \, m_{a_1} \, m_{a_2}
(v_a^i  - v_{a_1}^i )
(v_a^j  - v_{a_2}^j)  (m_a+m_{a_3})\cdots (m_a+m_{a_k})
\, , \nonumber \\
&& B_{(k)}^{ij} = { (2\, G)^{k+1}\over  R} \, \alpha^{-3k+2}\, 2^{-k} \, (-1)^{k+1}
\nonumber \\
&& {\sum_{a,a_1,a_2,\cdots, a_k\atop a_1,\cdots, a_k\ne a} } \prod_{\ell=1}^k |\vec v'_a
-\vec v'_{a_\ell}|^{-3} \, m_a'\, m_{a_1}' \, m_{a_2}' \, (v_a^{\prime i}  - v_{a_1}^{\prime i} )
(v_a^{\prime j}  - v_{a_2}^{\prime j} )  (m'_a+m'_{a_3})
\cdots (m'_a+m'_{a_k})
\, , \nonumber \\
&& \hskip 1in \hbox{for $k\ge 2$}\, .
\een
We shall use these results to compare the small-velocity limit of
\refb{e5.4} with the nonrelativistic results of
Section~\ref{s2}.

\subsection{Comparison of the first three terms}

We first compare the Newtonian limit of the first three terms of \refb{e5.4},
proportional to $(\nu\pm i\eps)^{-1}$, $\ln (\nu\pm i\eps)$ and
$(\nu\pm i\eps) (\ln (\nu\pm i\eps))^2$, with the corresponding terms of \refb{efourier}.
Substituting \refb{e3.10}--\refb{e3.15z} into \refb{e5.4}, we get
\ben \label{e4.17}
\tilde e^{ij}(\nu) &\simeq& {2 \, G \over R\, \alpha} \left[ {i\over \nu+i\eps} \sum_a m_a v_a^i v_a^j
- {i\over \nu-i\eps} \sum_a m'_a v_a^{\prime i} v_a^{\prime j}
\right] \nonumber \\
&-& {2 \, G  \over R\, \alpha}
\left[ \ln(\nu+i\eps) \sum_a m_a c_a^i v_a^j
- \ln(\nu-i\eps) \sum_a m'_a c_a^{\prime i} v_a^{\prime j}
\right] \nonumber \\
&-& { i\, G  \over R\, \alpha}
\Bigg[ (\nu+i\eps) (\ln(\nu+i\eps))^2 \sum_a m_a c_a^i c_a^j \nonumber \\*
&& \hskip 0.6in -(\nu-i\eps) (\ln(\nu-i\eps))^2 \sum_a m'_a c_a^{\prime i} c_a^{\prime j}
\Bigg]\, .
\een
Although the second line is not manifestly symmetric under $i\leftrightarrow j$, one can
verify using \refb{e1.28y} that it is. After symmetrizing in $i$ and $j$, \refb{e4.17}
agrees with \refb{efourier}.

\subsection{Two-to-two scattering}

We now consider scattering with two particles in both the initial and the final state.
Since we have already checked the expressions for $A_{(k)}^{ij}$ and
$B_{(k)}^{ij}$ for $k=0,1,2$, we focus on $k\ge 3$. For two particles, \refb{e3.16}
gives
\be\label{e3.20x}
A^{ij}_{(k)} = { 2\, G^{k+1}\over  R} \, (-1)^{k+1} \,
(m_1+m_2)^{k-1} \, v^{-3k}\, \alpha^{-3k+2}\, m_1 m_2\, v^i \, v^j\, ,
\ee
\be\label{e3.21x}
B^{ij}_{(k)} = { 2\, G^{k+1}\over  R} \, (-1)^{k+1} \,
(m'_1+m'_2)^{k-1} \, v^{\prime\, -3k}\, \alpha^{-3k+2}\, m'_1 m'_2\, v^{\prime i} v^{\prime j}\, .
\ee
Substituting these into \refb{e5.4}, using \refb{e3.10}, and keeping only the terms with
$k\ge 3$, we get
\ben\label{e5.4kk}
\tilde e^{ij}(\nu)|_{k\ge 3} &\simeq& - {2\, G\over R\alpha}\, \sum_{k\geq 3}{\nu^{k-1}\over k!}
\Bigg[ {m_1m_2\over m_1+m_2}\, \left( {G \, (m_1 +m_2) \over v^3 } \right)^k
i^{k-1}\, v^i \, v^j\, (\ln (\nu+i\eps))^k \nonumber \\*
&& \hskip 0.3in + {m'_1m'_2\over m'_1+m'_2}\, \left( {G \, (m_1' +m_2') \over v^{\prime 3} } \right)^k
(-i)^{k-1}\, v^{\prime i} \, v^{\prime j}\, (\ln (\nu-i\eps))^k \Bigg]\, .
\een
This agrees term by term with the $k\ge 3$ terms of \refb{e2.9xx}, to all orders. This is
an indication that the conjectured formul\ae\ \refb{eguess} and \refb{eguesstwo} are
correct for $2\to 2$ scattering.
We shall see in Section~\ref{sec-3pt}, however, that they fail in general for final
states with three or more particles, already at leading Newtonian order.

\subsection{Three-particle final state}
\label{sec-3pt}

Since a single counterexample suffices to show that \refb{eguess} and \refb{eguesstwo}
fail for three-particle final states, we can choose the masses, velocities and detector
direction conveniently. We take
\be
m_1=m_2=m_3=m, \quad \vec v_1=\vec 0, \quad \vec v_2=\vec v:= (v,0,0), \quad
\vec v_3 = -\vec v, \quad \vec n = (0,0,1)\, ,
\ee
and define
\be
\beta :={5\over 4}\, {G\, m\over v^3}\, .
\ee
Here $\vec v$ denotes the velocity of particle 2. This differs from its
use in Section~\ref{S:two_particle} for $\vec v_1-\vec v_2$, but should not cause any
confusion.
In this case, \refb{e1.28y} and \refb{eq:general-dr} give
\be
\vec c_1 = \vec 0, \quad \vec c_2 = \beta\,  \vec v, \quad
\vec c_3 = -\beta \vec v\, ,
\ee
\be
\vec d_{r,1} = \vec 0, \quad \vec d_{r,2} = (-1)^{r+1} {1\over r} \, \beta^{r+1}\,  \vec v, \quad
\vec d_{r,3} = -(-1)^{r+1} {1\over r} \, \beta^{r+1}\,  \vec v\, .
\ee
Substituting these into \refb{e1.25}, we find that the coefficient of the
$w^{-3} (\ln w)^2$ term in $e^{ij}$ is
\be
{G^4 m^4 \alpha^2\over R\, v^7}\times {125\over 16}\, \times \delta_{i1} \delta_{j1}\, .
\ee
Using \refb{e1.30} and \refb{e5.4}, this translates to
\be\label{eactual}
A_{(3)}^{ij} ={G^4 m^4 \over  R\, \alpha^7\, v^7}\times {125\over 16}\,
\times \delta_{i1} \delta_{j1}\, .
\ee

In the Newtonian limit, the conjectured formula of~\cite{2407.04128} takes the form
\refb{e3.16}. After some algebra, it gives, for this configuration,
\be
A_{(3);\rm conjectured}^{ij} ={G^4 m^4 \over R\, \alpha^7\, v^7}\times {225\over 16}\,
\times \delta_{i1} \delta_{j1}\, .
\ee
This differs from \refb{eactual}. The difference is proportional to
$\delta_{i1}\delta_{j1}$, which for $\vec n=(0,0,1)$ is transverse and has a nonzero
traceless part, so it cannot be removed by a gauge transformation. This shows that the
conjectured form \refb{eguess} and \refb{eguesstwo} of the coefficients
$A_{(k)}^{\mu\nu}$ and $B_{(k)}^{\mu\nu}$ is not valid in general when the initial or
final state contains three or more particles.

In Appendix~\ref{sb}, we describe another consistency test, which \refb{eguess} and
\refb{eguesstwo} fail for $k\ge 3$ but which our result \refb{e1.25} passes.

\sectiono{An all-orders conjecture for multiparticle relativistic scattering} \label{s5}

Using a combination of analytic computations and AI tools, we have obtained all-orders
expressions for the coefficients $A^{\mu\nu}_{(k)}$ and $B^{\mu\nu}_{(k)}$ for arbitrary
asymptotic velocities in general relativity, which we present in this section. Their
derivation relies on dropping certain terms that we have not been able to fully justify, so
the expressions given below should be regarded as a conjecture rather than a definitive
result.
Nevertheless, our analytic computation and the computation using AI tools rely on slightly
different assumptions---the former about the relevant integration regions in momentum
space, the latter about the relevant integration regions in position space---and they lead
to the same result. This agreement, together with the consistency checks described below,
gives us confidence in the conjecture.
We hope to present the full derivation, justifying these assumptions, in future
work~\cite{KKMSS:wip}.

We use the conventions of Section~\ref{S:comparisons}, with $\{p_a^\mu\}$ the outgoing
and $\{p_a^{\prime \mu}\}$ the incoming momenta. For each pair of outgoing particles
labeled by $a,b$ with $a\ne b$, we define
\begin{eqnarray}
D_{ab} &:=& \big[(p_a\cdot p_b)^2-p_a^2p_b^2\big]^{1/2}>0,\qquad
\sigma_{ab}:=-\frac{p_a\cdot p_b}{m_am_b}>1, \nonumber \\ %
\gamma_{ab} &:=& \frac{p_a\cdot p_b\big[\tfrac32p_a^2p_b^2-(p_a\cdot p_b)^2\big]}{D_{ab}^3}
=\frac{\sigma_{ab}(\sigma_{ab}^2-\tfrac32)}{(\sigma_{ab}^2-1)^{3/2}} ,
\end{eqnarray}
\begin{equation}\label{eq:Ca}
C_a^\mu:=\sum_{b\ne a}\Big[2G\gamma_{ab}\,p_b^\mu-\frac{Gp_a^2(p_b^2)^2}{D_{ab}^3}\,p_a^\mu\Big].
\end{equation}
Here $\gamma_{ab}=-\tau_{ab}$, with $\tau_{ab}$ defined in \refb{eq:tau-exact}.
For a momentum $p_b$, we define $A_b$ and record its contractions with $p_a$,
\begin{equation} \label{e54xyz}
\begin{gathered}
A_{b,\mu\nu}:=p_{b\mu}p_{b\nu}-\tfrac12 p_b^2\eta_{\mu\nu}\, ,\\
(A_b\cdot p_a)^\mu=(p_a\cdot p_b)p_b^\mu-\tfrac12p_b^2p_a^\mu,\qquad
p_a\cdot A_b\cdot p_a=(p_a\cdot p_b)^2-\tfrac12p_b^2p_a^2\, ,
\end{gathered}
\end{equation}
and for a four-vector $q$, we define
\begin{equation}
\mathcal D_b(q):=\big[(p_b\cdot q)^2-p_b^2\,q^2\big]^{1/2}.
\end{equation}
For incoming particles, we define similar quantities with $\{p_a\}$ replaced by
$\{p_a'\}$.

\def\zr{\check z}

We introduce a formal parameter $\zr$ and define
\begin{equation}
q^\mu_{ab}(\zr):=p_a^\mu+\zr(C_a^\mu-C_b^\mu)\, ,
\end{equation}
\begin{equation}\label{eq:force}
\begin{split}
\mathcal F^\mu_{ab}(\zr):=\frac{2G}{\mathcal D_b(q_{ab})^3}
\Big\{&2(A_b\cdot p_a)^\mu\big[(p_b\cdot q_{ab})(p_a\cdot p_b)-p_b^2(p_a\cdot q_{ab})\big]\\
&-(p_a\cdot A_b\cdot p_a)\big[(p_b\cdot q_{ab})p_b^\mu-p_b^2q_{ab}^\mu\big]\Big\}\, ,
\end{split}
\end{equation}
\begin{equation}
\mathcal F_a^\mu(\zr):=\sum_{b\ne a}\mathcal F^\mu_{ab}(\zr)\, ,
\end{equation}
with the sum over all other particles in the \emph{same} asymptotic set as $a$. All
factors of $q_{ab}$ appearing in \refb{eq:force} have the same argument $\zr$ as
$\mathcal F_a^\mu$. We denote by $\mathcal F^\mu_{a,r}$ the coefficient of $\zr^r$ in the
expansion of $\mathcal F^\mu_a(\zr)$ and define
\begin{equation}\label{eq:Ua}
\mathcal U_a^\mu(\zr):=-\sum_{r\ge 0}\frac{\mathcal F^\mu_{a,r}}{r+1}\,\zr^r=
-\int_0^1\mathcal F^\mu_a(s\zr)\,\dif s, \qquad \mathcal F^\mu_a(0) = -C_a^\mu\, .
\end{equation}
The coefficients $\mathcal F^\mu_{a,r}$, together with the asymptotic four-momenta,
determine the trajectories of the matter particles via
\be \label{ereltraj}
X_a^\mu(\lambda) = p_a^\mu \, \lambda + C_a^\mu \, \ln {\lambda\over L}
+ \sum_{r\ge 1} {\mathcal F^\mu_{a,r}\over r(r+1)} {(\ln{\lambda\over L})^r\over \lambda^r}
\, ,
\ee
where $\lambda$ is a parameter along the trajectory and $L$ is a length cutoff.
Analogous quantities are defined for the incoming particles.

\def\zt{\tilde z}

Next, we define
\begin{equation}
d_a:=-n\cdot p_a > 0\, ,
\end{equation}
\begin{equation}
\zeta_a(\zt):=\frac{\zt\,d_a}{1+\zt\,n\cdot C_a},
\end{equation}
\begin{equation}\label{eq:KX}
\mathcal K_X^{\mu\nu}(\zt):=2G\sum_a\frac1{1+\zt\,n\cdot C_a}
\Big[\Sym\big(p_a,\mathcal U_a(\zeta_a(\zt))\big)^{\mu\nu}
+\frac{p_a^\mu p_a^\nu}{d_a}\,n\cdot\mathcal U_a(\zeta_a(\zt))\Big]\, .
\end{equation}
Here
\begin{equation}
\Sym(A,B)^{\mu\nu} := (A^\mu B^\nu + A^\nu B^\mu)\, .
\end{equation}
The quantity $\mathcal K_X^{\mu\nu}(\zt)/R$ is the generating function for the
contributions to $A^{\mu\nu}_{(k)}$ from the outgoing matter part of the energy-momentum tensor. The
corresponding incoming contribution, which determines $B^{\mu\nu}_{(k)}$, is obtained by
replacing $p_a$ by $p'_a$.

The assumptions required to arrive at the expressions above are relatively mild. The main
assumption is that the gravitational field sourced by the gravitational stress-energy does
not affect the matter trajectories at the order considered.
The contribution to the waveform from the gravitational part of the
energy-momentum tensor requires additional assumptions about the relevant integration
regions, as noted at the beginning of this section.
We now construct this contribution, $\mathcal K_H^{\mu\nu}$.

Consider a superposition of two plane waves,
$H_{\mu\nu}=\varepsilon_a \wt A_{a,\mu\nu}e^{ik_a\cdot x}+\varepsilon_b
\wt A_{b,\mu\nu}e^{ik_b\cdot x}$, where
$\varepsilon_a,\varepsilon_b$ are bookkeeping parameters, $\wt A_a$, $\wt A_b$ are arbitrary
symmetric tensors, and $k_a$, $k_b$ are arbitrary four-vectors. We define
\begin{align}
\Gamma_{(1)\sigma\mu\nu}&:=\tfrac12(\partial_\mu H_{\sigma\nu}+\partial_\nu H_{\sigma\mu}-\partial_\sigma H_{\mu\nu}),\\
\Gamma^\rho_{(1)\mu\nu}&:=\eta^{\rho\sigma}\Gamma_{(1)\sigma\mu\nu},\qquad
\Gamma^\rho_{(2)\mu\nu}:=-H^{\rho\sigma}\Gamma_{(1)\sigma\mu\nu},\\
R_{(1)\mu\nu}&:=\partial_\rho\Gamma^\rho_{(1)\mu\nu}-\partial_\nu\Gamma^\rho_{(1)\mu\rho},\\
R_{(2)\mu\nu}&:=\partial_\rho\Gamma^\rho_{(2)\mu\nu}-\partial_\nu\Gamma^\rho_{(2)\mu\rho}
+\Gamma^\rho_{(1)\rho\lambda}\Gamma^\lambda_{(1)\mu\nu}-\Gamma^\rho_{(1)\nu\lambda}\Gamma^\lambda_{(1)\mu\rho},\\
\mathcal G_{(2)}^{\mu\nu}[H]&:=R_{(2)}^{\mu\nu}-H^{\mu\alpha}R_{(1)\alpha}{}^\nu-H^{\nu\alpha}R_{(1)\alpha}{}^\mu
-\tfrac12\eta^{\mu\nu}\big(\mathrm{tr}\,R_{(2)}-H^{\alpha\beta}R_{(1)\alpha\beta}\big)\notag\\
&\quad+\tfrac12H^{\mu\nu}\,\mathrm{tr}\,R_{(1)}+\tfrac12\,\mathrm{tr}\,H\Big(R_{(1)}^{\mu\nu}-\tfrac12\eta^{\mu\nu}\mathrm{tr}\,R_{(1)}\Big),\label{eq:G2}
\end{align}
where
\begin{equation}
\mathrm{tr}\,X :=\eta^{\mu\nu}X_{\mu\nu}\, .
\end{equation}
As usual, all indices are raised and lowered with $\eta$.
Up to a constant of proportionality, $\mathcal G_{(2)}^{\mu\nu}[H]$ is the energy-momentum
tensor of the gravitational field $H_{\mu\nu}$ at quadratic order.
An explicit expression for $\mathcal G^{\mu\nu}_{(2)}$ in terms of the metric perturbation can be found in~\cite[Eq.~(3.40)]{1912.06413}.
We now define $\mathcal V^{\mu\nu}[\wt A_a,k_a;\wt A_b,k_b]$ by
\begin{equation}\label{e5.19xx}
\mathcal G_{(2)}^{\mu\nu}[H]\big|_{\varepsilon_a\varepsilon_b}
=\mathcal V^{\mu\nu}[\wt A_a,k_a;\wt A_b,k_b]\,e^{i(k_a+k_b)\cdot x},
\end{equation}
where $\mathcal G_{(2)}^{\mu\nu}[H]\big|_{\varepsilon_a\varepsilon_b}$ is the
coefficient of $\varepsilon_a\varepsilon_b$ in $\mathcal G_{(2)}^{\mu\nu}[H]$.

To describe the gravitational fields produced by particles $a$ and $b$, we set
$\wt A_a=A_a$ and $\wt A_b=A_b$, with $A_b$ defined in \refb{e54xyz}. To compute the
waveform produced by the gravitational part of the energy-momentum tensor, we use the fact
that the time Fourier transform of the waveform is given by the full Fourier transform of
the energy-momentum tensor at momentum $\omega\, n$ (see {\it e.g.}
\cite{1804.09193,1912.06413}). Hence, to evaluate the contribution from \refb{e5.19xx}, we
set $k_a+k_b =\omega n$ and carry out a four-dimensional integral over $k_b$. Using various
approximations, we reduce this to a two-dimensional integral over the plane orthogonal to
$p_a$ and $p_b$ (see {\it e.g.} appendix A of \cite{1912.06413} for an example of such
manipulations). The radial integral produces the logarithm, while the angular integral is
carried out separately.
Since we have not yet been able to fully justify the approximations used in
this step, we do not describe it in detail and give only the final result. In this step, the
momentum integral runs from $1/u$ to the inverse size of the scattering region; the
exponential factor in \refb{e5.4} comes from momenta between $1/R$ and $1/u$.

We first consider the outgoing particles. For a pair of outgoing particles labeled by
$a,b$, let $e_1,e_2$ be \emph{any} fixed orthonormal spacelike basis of the
two-dimensional plane orthogonal to both $p_a$ and $p_b$. The construction below does not
depend on this choice, since it involves an average over $\phi$, which rotates $e_1,e_2$
within the plane. We define the projector onto that plane,
\be
\Pi_{ab}^{\mu\nu}:=e_1^\mu e_1^\nu+e_2^\mu e_2^\nu\, ,
\ee
and the unit vectors
\begin{equation}
e_\phi:=\cos\phi\,e_1+\sin\phi\,e_2, \qquad 0\le \phi \le 2\pi .
\end{equation}
We further define
\begin{equation}
\varpi^\mu_{ab}:=\frac{p_b\cdot n}{D_{ab}^2}\big[(p_a\cdot p_b)p_a^\mu-p_a^2p_b^\mu\big]\, ,
\ee
\begin{eqnarray}
&& k_a^\mu(\chi ):= e^\mu_\phi+\chi \,\varpi^\mu_{ab},\qquad k_b^\mu(\chi ):=-e^\mu_\phi
+\chi \,(n^\mu-\varpi^\mu_{ab}),
\nonumber \\
&& \Phi_{ab}^{\mu\nu}(\chi ;\phi):=\frac{\mathcal V^{\mu\nu}[A_a,k_a(\chi );A_b,k_b(\chi )]}
{\big(k_a(\chi )\cdot k_a(\chi )\big)\big(k_b(\chi )\cdot k_b(\chi )\big)} ,
\end{eqnarray}
where $\chi$ is a real number and $A_a$ and $A_b$ are defined in \refb{e54xyz}. Note that
\be
p_a\cdot\varpi_{ab}=0,\qquad  p_b\cdot\varpi_{ab}=p_b\cdot n\, .
\end{equation}
The function $\Phi_{ab}^{\mu\nu}(\chi ;\phi)$ is analytic in $\chi$ near $\chi=0$. In terms
of the vectors $C_a$ of \refb{eq:Ca} for the outgoing particles, we define
\begin{equation}
\beta_{ab}:=n\cdot C_b+\varpi_{ab}\cdot(C_a-C_b),\qquad b_\phi:=e_\phi\cdot(C_a-C_b).
\end{equation}
We also define the circle average
\begin{equation}
\langle f\rangle:=(2\pi)^{-1}\int_0^{2\pi}f(\phi)\,\dif\phi\, .
\end{equation}
We then define
\begin{equation}\label{eq:KH}
\mathcal K_H^{\mu\nu}(\zt):=-\sum_{a<b}\frac{16G^2}{D_{ab}}
\Big\langle\frac{\Phi_{ab}^{\mu\nu}(\chi _\phi;\phi)}{1+\zt\beta_{ab}}\Big\rangle,
\qquad \chi _\phi:=-\frac{\zt\,b_\phi}{1+\zt\beta_{ab}},
\end{equation}
where the sum runs over unordered pairs $a<b$ of outgoing particles. The quantity
$\mathcal K_H^{\mu\nu}(\zt)/R$ is the generating function for the contributions to
$A^{\mu\nu}_{(k)}$ from the gravitational part of the energy-momentum tensor; the
integration over $\phi$ arises from the angular integral mentioned above. The incoming
contribution is obtained by replacing $p_a$ by $p'_a$.

We now comment on how to carry out the circle average. We first expand
$\Phi^{\mu\nu}_{ab}(\chi ;\phi)$ in a power series in $\chi$ at fixed $\phi$,
$\Phi^{\mu\nu}_{ab}=\sum_{j\ge0}N^{\mu\nu}_{j,ab}(\phi)\,\chi^j$. Each
$N^{\mu\nu}_{j,ab}(\phi)$ is a polynomial of degree at most $j+2$ in $\cos\phi$ and
$\sin\phi$, the components of $e_\phi$ in the basis $(e_1,e_2)$. To average any such
polynomial, we can expand it into monomials $\cos^a\phi\sin^b\phi$ (equivalently, contract
the $e_\phi$'s using $\Pi_{ab}$) and use
\begin{equation}\label{eq:moments}
\big\langle\cos^a\phi\,\sin^b\phi\big\rangle=
\begin{cases}
\dfrac{(a-1)!!\,(b-1)!!}{(a+b)!!}, & a,b\text{ both even},\\[2mm]
0, & \text{otherwise},
\end{cases}
\qquad (-1)!!:=1\, .
\end{equation}
This turns \refb{eq:KH}, order by order in $\zt$, into an algebraic expression in the
momenta and $n$.

We are now ready to state the conjectured expressions for $A_{(k)}^{\mu\nu}$ and
$B_{(k)}^{\mu\nu}$. We first define
\begin{equation}
\mathcal K^{\mu\nu}(\zt):=\mathcal K_X^{\mu\nu}(\zt)+\mathcal K_H^{\mu\nu}(\zt)\, .
\end{equation}
Let $\mathcal K_{(r)}^{\mu\nu}$ be the coefficient of $\zt{}^r$ in the expansion of
$\mathcal K^{\mu\nu}(\zt)$. Our conjecture is that, for $k\ge1$,
\begin{equation}\label{eq:W}
A^{\mu\nu}_{(k)}= {1\over R} \, \mathcal K_{(k-1)}^{\mu\nu}[p],\qquad
B^{\mu\nu}_{(k)}={1\over R}\, \mathcal K_{(k-1)}^{\mu\nu}[p'],
\end{equation}
where $[p]$ and $[p']$ indicate that $\mathcal K$ is evaluated on the outgoing and incoming
momenta, respectively. Once $A_{(k)}^{\mu\nu}$ and $B_{(k)}^{\mu\nu}$ have been
determined, we can use \refb{e5.4} and \refb{e1.30} to determine the late- and early-time
waveforms. As an illustration, Appendix~\ref{sd} gives the resulting explicit expression
for $A_{(3)}^{\mu\nu}$.

\paragraph{Consistency checks.}
We have carried out the following consistency checks of the conjecture:
\begin{enumerate}
\item We have checked the harmonic gauge condition
\be
n_\mu A_{(k)}^{\mu\nu}=0, \qquad n_\mu B_{(k)}^{\mu\nu}=0\, ,
\ee
which is also a test of the conservation of the total energy-momentum tensor.
\item We have checked that, in the Newtonian limit, the conjecture reproduces our results
\refb{e1.25}, \refb{e1.28y} and \refb{eq:general-dr}.
\item We have checked that the conjecture passes the subdivision test of
Appendix~\ref{sb}: with the prescription described there, the result is unchanged when a
particle of momentum $p_a$ is replaced by two particles of momenta $f\, p_a$ and
$(1-f)\, p_a$, with $0<f<1$.
\item For two-to-two scattering, the conjecture agrees with \refb{eguess} and
\refb{eguesstwo}, for which there is independent evidence~\cite{2407.04128}. Indeed, in
this case $C_1$ and $C_2$ lie in the plane spanned by $p_1$ and $p_2$, so
$b_\phi=e_\phi\cdot(C_1-C_2)=0$, and hence $\chi_\phi=0$. Thus, the entire $\zt$
dependence of $\mathcal K_H^{\mu\nu}(\zt)$ in \refb{eq:KH} comes from the denominator
$1+\zt\beta_{12}$.
In $\mathcal K_X^{\mu\nu}(\zt)$, defined in \refb{eq:KX}, the prefactors
$1/(1+\zt\,n\cdot C_a)$ combine with $\mathcal U_a(\zeta_a(\zt))$ to give the same overall
factor $1/(1+\zt\beta_{12})$. The
entire $\zt$ dependence of both $\mathcal K_X^{\mu\nu}$ and $\mathcal K_H^{\mu\nu}$
therefore comes from a simple pole at $\zt=-1/\beta_{12}$, where, for two-to-two
scattering, $\beta_{12}=-2G\gamma_{12}(d_1+d_2)$; in the Newtonian limit, $\beta_{12}$
reduces to $\gamma$ of \refb{e2.3z}. The Taylor expansion in $\zt$ then reproduces
\refb{eguess} and \refb{eguesstwo}.
\item In the probe limit, where one particle carries a much larger four-momentum than the
others, the conjecture again agrees with \refb{eguess} and \refb{eguesstwo}, which are
independently supported in this limit~\cite{2408.07329}.
\item The conjecture has also been shown to agree with the next-to-leading, $O(\alpha^3)$,
terms in the small-velocity expansion of the highest-logarithmic waveform
coefficients~\cite{unpublished}.
\end{enumerate}

\paragraph{Further consequences of the conjecture.}
We have also examined some consequences of the conjecture:
\begin{enumerate}
\item For a process with a single massive object in both the initial and final states,
together with any number of massless particles in the final state, both the late- and
early-time tails, i.e.\ the terms $u^{-k}(\ln|u|)^{k-1}$ with $k\ge1$, vanish,
once the contributions of the exponential factors in \refb{e5.4} are
included, leaving only the memory term. This is the situation relevant for binary black hole mergers, since the
bound binary can be treated as a single object for this purpose. This appears to be
consistent with~\cite{Damour:2007yf,Ivanov:2025ozg} for gravitational-wave emission from
compact binaries during the inspiral phase.
\item If the final state contains several massive and several massless particles, then the
result depends on the momenta of the individual massless particles, and not just on their
total momentum. This is in contrast with the coefficients of the $u^{-1}$ and
$u^{-2}\ln|u|$ terms, which depend only on the total momentum of the massless particles.
For these two coefficients, there are general arguments explaining why the dependence on
the individual massless momenta drops out~\cite{2603.09844,2603.22681,2508.17919}; to the
best of our knowledge, these arguments have not been extended to higher orders. Thus, as
far as we know, our conjecture does not contradict any established result.
\end{enumerate}

\paragraph{Acknowledgements.}
We would like to thank Gianni Boschetti, Miguel Campiglia and Alok Laddha
for useful discussions.
This work was supported by the
Department of Atomic Energy, Government of India, under project no.~RTI4019.
R.M. thanks the Leinweber Institute for Theoretical Physics at Stanford for its
hospitality during the completion of part of this work.
The work of A.S. was supported by the ICTS-Infosys Madhava
Chair Professorship.

\paragraph{Statement on AI use.}
The analysis presented in this paper was developed with assistance from AI tools:
ChatGPT (OpenAI), Claude (Anthropic) and an internal model at Google DeepMind.
AI tools were also used to copy-edit the manuscript. The manuscript was written by the
authors, who take full responsibility for its content, including any errors.

\appendix

\sectiono{Analysis of the error terms} \label{sa}

In this appendix, we estimate the error terms $\EE_1$, $\EE_2$ and $\EE_3$, defined in
\refb{e1.9}, \refb{e1.17} and \refb{e1.25xxy}, respectively, and show that they do not
contribute at the order retained in Section~\ref{s2}.

\subsection{Analysis of $\EE_1$} \label{sa.1}

We show directly from the definition \refb{e1.9} that $\EE_1^{ij}=O(\alpha^3)$ in the
small-$\alpha$ expansion. Expanding the integrand of \refb{e1.9} in the retardation
shift $\vec n\cdot\vec x$ and separating the first-order term gives
\be\label{eq:a1-first-retardation}
\EE_1^{ij}={2G\over R}n_k\p_u S^{kij}+\mathcal R_{\ge2}^{ij},
\qquad S^{kij}:=\int d^3x\,x^kT^{ij}(u,\vec x).
\ee
Here $\mathcal R_{\ge2}^{ij}$ denotes terms of second and higher order in the retardation
shift; we estimate their $\alpha$ order below.
Since the gravitational stress-energy falls off only as $T^{ij}=O(r^{-4})$,
the integral defining $S^{kij}$ is logarithmically divergent by power counting; it
converges only because the leading, $O(r^{-4})$, part of $T^{ij}$ at large $r$ is even
under parity, while $x^k$ is odd. In any case, that part is $u$-independent and does not
contribute to $\p_uS^{kij}$ in \refb{eq:a1-first-retardation}.

To express this moment of $T^{ij}$ in
terms of moments of $T^{00}$ and $T^{0i}$ using conservation, as in \refb{e1.13}, we
define the mass octupole moment and the first spatial moment of the angular-momentum
density, respectively,
\begin{align}
I^{kij}&:=\int d^3x\,x^kx^ix^jT^{00}(u,\vec x),\label{eq:a1-mass-moment}\\
J^{ki|j}&:=\int d^3x\,x^j\bigl(x^kT^{0i}-x^iT^{0k}\bigr)(u,\vec x).\label{eq:a1-current-moment}
\end{align}
Since the total $T^{\mu\nu}$ is symmetric and conserved, integrating by parts as in
\refb{e1.13} yields
\be\label{eq:a1-moment-identities}
\p_u^2 I^{kij}=2\bigl(S^{kij}+S^{ikj}+S^{jki}\bigr),
\qquad \p_u J^{ki|j}=S^{kij}-S^{ikj}.
\ee

As noted after \refb{eq:a1-first-retardation}, the gravitational
stress-energy falls off only as $T^{\mu\nu}=O(r^{-4})$. As a result, the integrals
defining $I^{kij}$ and $J^{ki|j}$, like the moment in \refb{e1.13}, do not converge at
large $r$, and \refb{eq:a1-moment-identities} should be understood with a large-$r$
cutoff $\Lambda$. With the cutoff in place, the surface terms generated by the
integrations by parts vanish as $\Lambda\to\infty$; for the leading, $O(r^{-4})$, part of
$T^{\mu\nu}$ this follows from parity. The moments $I^{kij}$ and $J^{ki|j}$ diverge
linearly in $\Lambda$, but only through the parts of the far field fixed by the total
mass $M$, the total momentum $\vec P$ and the mass dipole
$\vec D(u):=\sum_a m_a\vec X_a(u)$: the divergent part of $I^{kij}$ is proportional to
$M\vec D(u)$, and that of $J^{ki|j}$ to $M\vec P$. Since $\dot{\vec D}=\vec P$ is
conserved, these divergent parts are annihilated by $\p_u^2$ and $\p_u$, respectively, so
\refb{eq:a1-moment-identities} holds with the $u$-derivatives taken before the cutoff is
removed.

Solving \refb{eq:a1-moment-identities}, together with its $i\leftrightarrow j$
counterpart, for $S^{kij}$ and substituting into \refb{eq:a1-first-retardation}, we
obtain
\be\label{eq:a1-retardation-moments}
\EE_1^{ij}={G\over3R}n_k\p_u^3 I^{kij}
+{2G\over3R}n_k\p_u^2\bigl(J^{ki|j}+J^{kj|i}\bigr)+\mathcal R_{\ge2}^{ij}.
\ee
Unlike in \refb{e1.13}, conservation does not reduce this correction to moments of
$T^{00}$ alone. However, both moments can be expressed in terms of the particle
trajectories at leading Newtonian order, as for the mass quadrupole in \refb{eq:quad-particle}:
\be\label{eq:a1-particle-mass-moment}
I^{kij}\simeq\sum_a m_aX_a^kX_a^iX_a^j,
\ee
\be\label{eq:a1-particle-current-moment}
J^{ki|j}\simeq\sum_a m_aX_a^j
\bigl(X_a^k\dot X_a^i-X_a^i\dot X_a^k\bigr),
\qquad \dot X_a^i:=\p_uX_a^i.
\ee
Note that conservation has been applied to the total source before making these
particle approximations.

Equations \refb{e1.20}--\refb{e1.23}, together with
$\p_u=\alpha^3\p_w$ from \refb{e1.18}, give the $\alpha$ orders
\be\label{eq:a1-moment-scaling}
X_a^i=O(\alpha^{-2}),\qquad \dot X_a^i=O(\alpha),\qquad
I^{kij}=O(\alpha^{-6}),\qquad J^{ki|j}=O(\alpha^{-3}).
\ee
Hence the differentiated moments in \refb{eq:a1-retardation-moments} satisfy
\be\label{eq:a1-derived-orders}
\p_u^3 I^{kij}=O(\alpha^9\alpha^{-6})=O(\alpha^3),
\qquad
\p_u^2 J^{ki|j}=O(\alpha^6\alpha^{-3})=O(\alpha^3).
\ee
Within this formal source-moment expansion, each additional retardation insertion
$x^k\p_u$ brings one further power of $\alpha$, since $x^k=O(\alpha^{-2})$ and
$\p_u=O(\alpha^3)$, giving $\mathcal R_{\ge2}^{ij}=O(\alpha^4)$.
Thus, at each retained asymptotic order,
\be\label{eq:a1-retardation-scaling}
\EE_1^{ij}=O(\alpha^3),
\ee
which is subleading to the $O(\alpha^2)$ waveform in \refb{e1.25}.

\subsection{Analysis of $\EE_2$} \label{sa.2}

In this subsection, we show that the correction term $\EE_2$ defined in \refb{e1.17},
which captures the effects of the kinetic energy of the matter particles and of the
Newtonian gravitational energy, does not contribute at the order retained in
Section~\ref{s2}.

\subsubsection{Kinetic energy of the matter particles}

The kinetic energy of the matter particles multiplies each term $m_aX_a^iX_a^j$ in
\refb{eq:quad-particle} by the relative factor\footnote{The matter part of $T^{00}$, omitting gravitational-potential corrections, is
$T^{00}_{\mathrm{matter}}(u,\vec x)=\sum_a m_a(1-|\dot{\vec X}_a|^2)^{-1/2}
\delta^{(3)}(\vec x-\vec X_a(u))$.
Expanding the energy factor gives $m_a[1+\frac12|\dot{\vec X}_a|^2+\cdots]$;
after subtracting the rest-mass term in \refb{e1.17}, the kinetic factor
multiplies $m_aX_a^iX_a^j$ inside the two $u$ derivatives.}
 \be
 {1\over 2} \left({d \vec X_a\over du}\right)^2 =
 {1\over 2} \alpha^6 \left({d \vec X_a\over dw}\right)^2
 ={1\over 2} \alpha^ 2\left[\vec v_a + \vec c_a\, w^{-1}
- w^{-1} \sum_{r\ge 1}
\vec d_{r,a}  r z^r + \OO(w^{-2})\right]^2\, .
 \ee
Since this factor is $O(\alpha^2)$, the corresponding contribution to $e^{ij}$ is
$O(\alpha^4)$ and can be neglected.

\subsubsection{Gravitational potential energy}

The gravitational potential energy contributes through
$\int d^3x\,x^ix^jT^{00}_{\rm grav}$. Its large-$r$ tail is $u$-independent and is removed
by the two $u$-derivatives in \refb{e1.17}, as explained in footnote~\ref{fn:quad-div}.
The remaining contribution comes from distances of the order of the particle separations,
so we can estimate it by replacing $x^ix^j$ by $X^iX^j$ for a typical particle and
multiplying by the mutual gravitational interaction energy of the particles. This gives
a contribution to $e^{ij}$ of order
 \be
 {G\over R} {d^2\over d u^2}  \left[X^i X^j \sum_{a<b} m_a m_b
 {G\, \over |\vec X_a - \vec X_b|}\right]\, .
 \ee
Comparing this with \refb{eq:quad-particle}, we see that the relative factor is of order
 \be
G\, m\,  {1\over |\vec X_a-\vec X_b|} \sim G\, m\,
 {\alpha^2 \over |\vec v_a -\vec v_b| \, w}\, ,
 \ee
where we have used \refb{e1.20}. Since this factor is $O(\alpha^2)$, this contribution is
also subleading.

 \subsection{Analysis of $\EE_3$} \label{sa.3}

The first term on the right-hand side of \refb{e1.25xxy} accounts for the gravitational
force on the $a$'th particle due to the field produced by the rest masses of the other
particles. However, the gravitational field is also sourced by the gravitational field
energy and by the kinetic energies of the particles, and these additional sources
contribute to $\EE_3$. As shown in Appendix~\ref{sa.2}, they are suppressed by a factor of
$\alpha^2$ relative to the rest-mass energy. Hence, we expect the corresponding forces to be
suppressed by $\alpha^2$ relative to the Newtonian force, so that they change $\vec X_a$
only at $O(\alpha^0)$, beyond the $O(\alpha^{-2})$ terms retained in Section~\ref{s2}.

\sectiono{A subdivision test of the all-orders conjecture} \label{sb}

In this appendix, we propose another consistency test of the formul\ae\ \refb{eguess} and
\refb{eguesstwo}, conjectured in~\cite{2407.04128}. We formally subdivide the $c$'th outgoing particle into two coincident
pieces, with momenta and masses
\be
p_{c'}=f  p_c,\qquad m_{c'}=f  m_c,\qquad
p_{c''}=(1-f )p_c,\qquad m_{c''}=(1-f )m_c,
\qquad 0<f <1.
\ee
The mass scaling follows from $p^2=-m^2$. We assign both pieces the
original worldline, so their matter energy-momentum tensors sum to the
original one. In this formal subdivision, mutual interactions between
the two pieces are excluded as self-interactions of the original source;
in particular, terms involving the singular factor $\tau_{c'c''}$ are
omitted. This prescription is not a limit of scattering with distinct
asymptotic velocities.

The exact definition \refb{eq:tau-exact} is invariant under simultaneous
positive rescaling of either momentum and its mass: under
$(p_b,m_b)\mapsto(f  p_b,f  m_b)$, its numerator and denominator
both scale as $f ^3$. Thus, for each unsplit particle $a$ whose
velocity differs from that of particle $c$,
\be
\tau_{ac'}=\tau_{ac''}=\tau_{ac},\qquad
\tau_{c'a}=\tau_{c''a}=\tau_{ca}.
\ee
These identities are exact and require no nonrelativistic approximation. Consequently, for
$k\le 2$, the summands in \refb{eamunudef}--\refb{eguess} are homogeneous of degree one in
each momentum they contain, with the corresponding mass scaled simultaneously. Replacing
$p_c$ in any one slot by the pair $p_{c'}=f\,p_c$ and $p_{c''}=(1-f)\,p_c$ and summing
therefore reproduces the original term, since $f+(1-f)=1$. The same holds when $c$ appears
in both $a_1$ and $a_2$ in \refb{eguess} with $k=2$, since the summand is then linear in
each of the two slots. Hence $A^{\mu\nu}_{(k)}$ with $k\le 2$ is invariant under the
subdivision, and the same argument applies to incoming particles and
$B^{\mu\nu}_{(k)}$.

This property, however, is violated by the conjectured formula \refb{eguess} for $k\ge 3$.
For example, consider $A^{\mu\nu}_{(3)}$ and sum over $a_3=c',c''$ with fixed
$a,a_1,a_2\ne c$. After the sum, the factor $n\cdot(p_a+p_c)$ is replaced by
$n\cdot(p_a+p_{c'})+n\cdot(p_a+p_{c''})=n\cdot(2p_a+p_c)$, which differs from the original
factor $n\cdot(p_a+p_c)$. This shows that the conjectured formul\ae\ \refb{eguess} and
\refb{eguesstwo} do not satisfy this formal subdivision test.

Under the same prescription, our Newtonian formula \refb{e1.25} is unchanged.
The forces exerted by the other particles on the two coincident pieces give
them identical trajectories, and since their mutual interaction is excluded, their
trajectory coefficients coincide with those of the original particle, while their masses
add to the original mass:
\be
m_c=m_{c'}+m_{c''},\qquad
\vec c_c=\vec c_{c'}=\vec c_{c''},\qquad
\vec d_{r,c}=\vec d_{r,c'}=\vec d_{r,c''}\quad(r\ge1).
\ee
Since \refb{e1.25} is linear in $m_a$, the contribution of the original
particle is therefore recovered by summing the contributions of its two pieces.

\sectiono{Extraction of the highest-logarithmic coefficients}\label{sc}

The formal $w,z$ counting in Section~\ref{s2} selects the highest
logarithm at each inverse-time order without computing the lower
logarithms. Such lower logarithms need not vanish: the full gravitational
waveform contains, for example, a spin-dependent $u^{-2}$ contribution
\cite{2106.10741}, and the distinction between retained and lower
logarithms also appears in the electromagnetic analysis of
\cite{2501.07328}. To state the coefficient extraction precisely,
suppose that the leading Newtonian spatial waveform admits a
differentiable asymptotic expansion
\be\label{eq:waveform-log-expansion}
e^{ij}(w)\sim E_\infty^{ij}
+\sum_{k=1}^{\infty}C_k^{ij}\frac{(\ln w)^{k-1}}{w^k}
+\sum_{k=2}^{\infty}\sum_{r=0}^{k-2}H_{k,r}^{ij}
\frac{(\ln w)^r}{w^k},\qquad w\to+\infty\, .
\ee
The coefficients are independent of $w$: $E_\infty^{ij}$ is the
outgoing asymptotic value, $C_k^{ij}$ are the highest-logarithmic
coefficients computed in the main text, and $H_{k,r}^{ij}$ collect the
lower logarithms left undetermined. Since
$w^{-k}(\ln w)^{k-1}=w^{-1}z^{k-1}$, the coefficients $E_\infty^{ij}$ and $C_k^{ij}$ are
precisely those retained by the approximation in Section~\ref{s2}.

Each $C_k^{ij}$ can be isolated directly from the waveform.
With $D:=w\p_w$, for any fixed $k\ge1$,
\be\label{eq:waveform-log-projector}
C_{k}^{ij}=\frac{(-1)^{1+k(k-1)/2}}{k!\prod_{\ell=1}^{k-1}\ell!}
\lim_{w\to\infty}w^{k}D(D+k)^{k-1}
\prod_{\ell=1}^{k-1}(D+\ell)^\ell e^{ij}(w)\, ,
\ee
where, for $k=1$, the empty products are understood to be one.
This follows from the identity
$(D+m)\,w^{-m}(\ln w)^p=p\,w^{-m}(\ln w)^{p-1}$. The factors $(D+\ell)^\ell$ annihilate
the terms at $w^{-\ell}$ for $1\le\ell\le k-1$, since their logarithmic
degree is at most $\ell-1$. The factor $(D+k)^{k-1}$ removes the lower
logarithms at the target order, while $D$ annihilates $E_\infty^{ij}$;
acting on the remaining term $C_k^{ij}(\ln w)^{k-1}/w^k$, the operator produces the
inverse of the prefactor in \refb{eq:waveform-log-projector}.
The extraction is valid provided the asymptotic remainder remains
subleading after the indicated differentiations; convergence of the
infinite series is not required.

\sectiono{Explicit expression for $A_{(3)}^{\mu\nu}$} \label{sd}

In this appendix, we give the explicit expression for $A_{(3)}^{\mu\nu}$ computed from
\refb{eq:W}. We use the following quantities, some of which were introduced in
Sections~\ref{S:comparisons} and~\ref{s5}:
\begin{equation}
\begin{gathered}
m_a:=\sqrt{-p_a^2},\qquad d_a:=-n\cdot p_a,\qquad \delta_a:=\frac{d_a}{m_a},\\
\sigma_{ab}:=-\frac{p_a\cdot p_b}{m_am_b},\qquad
\tau_{ab}=-\frac{\sigma_{ab}(2\sigma_{ab}^2-3)}{2(\sigma_{ab}^2-1)^{3/2}} \, .
\end{gathered}
\end{equation}
We further define
\begin{equation}
\widehat C_a^{\,\mu}:=\frac{2G}{p_a\cdot n}
\sum_{c\neq a}\tau_{ac}\,n_\rho\,p^{\mu}_{[a}p^{\rho}_{c]}\, ,
\qquad n\cdot\widehat C_a=0 \, ,
\end{equation}
where we use the convention
\be
P_{[a}Q_{b]}:=P_a Q_b-P_b Q_a, \qquad X^{(\mu}Y^{\nu)} := {1\over 2} (X^\mu Y^\nu + X^\nu Y^\mu)\, .
\ee
For any pair of outgoing particles $a,b$, we also define
\begin{gather}
\mathcal Z_{ab}:=\widehat C_a-\widehat C_b,\qquad
w^\mu_{ab}:= d_b\,p_a^\mu-d_a\,p_b^\mu=n_\rho\,p^{\mu}_{[b}p^{\rho}_{a]},\\
\zeta_{ab}:=\frac{p_a\cdot\mathcal Z_{ab}}{m_a},\qquad \zeta_{ba}:= -\frac{p_b\cdot\mathcal Z_{ab}}{m_b},
\qquad  \kappa_{ab}:=2\sigma_{ab}^4-6\sigma_{ab}^2+1 ,\\
\mathcal Z_{ab\perp}:=\mathcal Z_{ab}-\frac{(\zeta_{ab}+ \sigma_{ab} \zeta_{ba})\,p_a/m_a-(\zeta_{ba}+\sigma_{ab}\zeta_{ab})\,p_b/m_b}{\sigma_{ab}^2-1}\, .
\end{gather}
In terms of these quantities, $A^{\mu\nu}_{(3)}$ is given by
\begin{equation}
A^{\mu\nu}_{(3)}=\frac1R\sum_{a<b}\Big[\mathcal T^{\mu\nu}_{ab}+n^{(\mu}\xi^{\nu)}_{ab}-\tfrac12\eta^{\mu\nu}\,n\cdot\xi_{ab}\Big],
\end{equation}
where the physical (transverse) part is
\begin{align}
\mathcal T^{\mu\nu}_{ab}=\frac{G^2}{(\sigma_{ab}^2-1)^{7/2}}\bigg\{
&-\frac{1}{m_am_b}\Big[(2\sigma_{ab}^2+3)(\zeta_{ab}^2+\zeta_{ba}^2)+10\,
\sigma_{ab}\,\zeta_{ab} \zeta_{ba}\nonumber\\
&\qquad\qquad-\tfrac13(\sigma_{ab}^2-1)(4\sigma_{ab}^4-12\sigma_{ab}^2+11)\,\mathcal Z_{ab}^2\Big]\,w_{ab}^\mu w_{ab}^\nu\nonumber\\
&+\tfrac43(\sigma_{ab}^2-1)\Big[(3\sigma_{ab}\delta_a-\kappa_{ab}\,\delta_b)\,\zeta_{ab}+
(3\sigma_{ab}\delta_b
- \kappa_{ab}\,\delta_a) \, \zeta_{ba}\Big]\,
w_{ab}^{(\mu}\mathcal Z_{ab}^{\nu)}\nonumber\\
&-\tfrac23(\sigma_{ab}^2-1)^2(2\sigma_{ab}^2-1)\,m_am_b\,(\delta_a^2+\sigma_{ab}\delta_a\delta_b+\delta_b^2)\,
\mathcal Z_{ab}^\mu\mathcal Z_{ab}^\nu\bigg\},
\end{align}
and the remaining terms are pure gauge, of the form
$n^{(\mu}\xi^{\nu)}-\frac12\eta^{\mu\nu}n\cdot\xi$, with
\begin{align}
\xi^\mu_{ab}=\frac{G^2}{(\sigma_{ab}^2-1)^{3/2}}\bigg\{&\frac{14\sigma_{ab}^2-13}{3}\,\mathcal Z_{ab\perp}^2
\Big[(\delta_a-\sigma_{ab}\delta_b)\,m_b\,p_a^\mu-(\sigma_{ab}\delta_a-\delta_b)\,m_a\,p_b^\mu\Big]\nonumber\\
&+\tfrac23(\sigma_{ab}^2-1)(2\sigma_{ab}^2-1)\,m_am_b\Big[(n\cdot\mathcal Z_{ab\perp})\,\mathcal Z_{ab\perp}^\mu+\tfrac52\,\mathcal Z_{ab\perp}^2\,n^\mu\Big]\bigg\}.
\end{align}
The contribution of each pair separately satisfies the transversality condition
$n_\mu(\cdots)^{\mu\nu}=0$. Only $\mathcal T_{ab}$ carries physical content.

\bibliographystyle{apsrev4-1long}
\bibliography{newtonian_references}

\end{document}